\documentclass[submission, PhysCore]{SciPost}

\hypersetup{
    colorlinks,
    linkcolor={red!50!black},
    citecolor={blue!50!black},
    urlcolor={blue!80!black}
}

\usepackage[bitstream-charter]{mathdesign}
\usepackage{caption,subcaption}
\usepackage{comment}
\DeclareSymbolFont{usualmathcal}{OMS}{cmsy}{m}{n}
\DeclareSymbolFontAlphabet{\mathcal}{usualmathcal}
\usepackage{siunitx}
\usepackage{upgreek}

\begin{document}

\begin{center}{\Large \textbf{
Diffusion of muonic hydrogen in hydrogen gas and the measurement of the 1$s$ hyperfine splitting of muonic hydrogen
}}\end{center}

\begin{center}
J.~Nuber\textsuperscript{1,2$\star$},
A.~Adamczak\textsuperscript{3},
M.~Abdou~Ahmed\textsuperscript{4},
L.~Affolter\textsuperscript{2},
F.~D.~Amaro\textsuperscript{5},
P.~Amaro\textsuperscript{6},
P.~Carvalho\textsuperscript{6},
Y.-H.~Chang\textsuperscript{7},
T.-L.~Chen\textsuperscript{7},
W.-L.~Chen\textsuperscript{7},
L.~M.~P.~Fernandes\textsuperscript{5},
M.~Ferro\textsuperscript{6},
D.~Goeldi\textsuperscript{2},
T.~Graf\textsuperscript{4},
M.~Guerra\textsuperscript{6},
T.~W.~H\"{a}nsch\textsuperscript{8,9}, 
C.~A.~O.~Henriques\textsuperscript{5},
M.~Hildebrandt\textsuperscript{1},
P.~Indelicato\textsuperscript{10},
O.~Kara\textsuperscript{2},
K.~Kirch\textsuperscript{1,2},
A.~Knecht\textsuperscript{1},
F.~Kottmann\textsuperscript{1,2},
Y.-W.~Liu\textsuperscript{7},
J.~Machado\textsuperscript{6},
M.~Marszalek\textsuperscript{1,2},
R.~D.~P.~Mano\textsuperscript{5},
C.~M.~B.~Monteiro\textsuperscript{5},
F.~Nez\textsuperscript{10},
A.~Ouf\textsuperscript{11},
N.~Paul\textsuperscript{10},
R.~Pohl\textsuperscript{11,12},
E.~Rapisarda\textsuperscript{1},
J.~M.~F.~dos~Santos\textsuperscript{5}, 
J.~P.~Santos\textsuperscript{6},
P.~A.~O.~C.~Silva\textsuperscript{5},
L.~Sinkunaite\textsuperscript{1,2},
J.-T.~Shy\textsuperscript{7},
K.~Schuhmann\textsuperscript{2},
S.~Rajamohanan\textsuperscript{11},
A.~Soter\textsuperscript{2},
L.~Sustelo\textsuperscript{6},
D.~Taqqu\textsuperscript{1,2},
L.-B.~Wang\textsuperscript{7},
F.~Wauters\textsuperscript{12,13},
P.~Yzombard\textsuperscript{10},
M.~Zeyen\textsuperscript{2},
J.~Zhang\textsuperscript{2},
A.~Antognini\textsuperscript{1,2$\dagger$}
\end{center}

\begin{center}
{\bf 1} Paul Scherrer Institute, 5232 Villigen–PSI, Switzerland
\\
{\bf 2} Institute for Particle Physics and Astrophysics, ETH Zurich, 8093 Zurich, Switzerland
\\
{\bf 3} Institute of Nuclear Physics, Polish Academy of Sciences, PL-31342 Krakow, Poland
\\
{\bf 4} Institut f\"ur Strahlwerkzeuge, Universit\"at Stuttgart, 70569 Stuttgart, Germany
\\
{\bf 5} LIBPhys-UC, Department of Physics, University of Coimbra, \mbox{P-3004-516 Coimbra, Portugal}
\\
{\bf 6} Laboratory of Instrumentation, Biomedical Engineering and Radiation Physics (LIBPhys-UNL), Department of Physics, NOVA School of Science and Technology, NOVA University Lisbon, 2829-516 Caparica, Portugal
\\
{\bf 7} LPhysics Department, National Tsing Hua University, Hsincho 300, Taiwan
\\
{\bf 8} Ludwig-Maximilians-Universit\"at, Fakult\"at f\"ur Physik, 80799 Munich, Germany
\\
{\bf 9} Max Planck Institute of Quantum Optics, 85748 Garching, Germany
\\
{\bf 10} Laboratoire Kastler Brossel, Sorbonne Universit\'{e}, CNRS, ENS-Universit\'e PSL, Coll\`{e}ge de France, 75005 Paris, France
\\
{\bf 11} QUANTUM, Institute of Physics, Johannes Gutenberg-Universit\"at Mainz, 55099 Mainz, Germany
\\
{\bf 12} Excellence Cluster PRISMA+, Johannes Gutenberg-Universit\"at Mainz, 55099 Mainz, Germany
\\
{\bf 13} Institute of Nuclear Physics, Johannes Gutenberg-Universit\"at Mainz, 55099 Mainz, Germany
\\

${}^\star${\small \sf jonas.nuber@psi.ch}
${}^\dagger${\small \sf aldo.antognini@psi.ch}
\end{center}

\begin{center}
\today
\end{center}


\section*{Abstract}
{\bf
The CREMA collaboration is pursuing a measurement of the ground-state hyperfine splitting (HFS) in muonic hydrogen ($\upmu$p)  with 1~ppm accuracy by means of pulsed laser spectroscopy. 
In the proposed experiment, the $\upmu$p atom is excited by a laser pulse from the singlet  to the triplet hyperfine sub-levels, and is quenched back to the singlet state by an inelastic collision with a H$_2$ molecule. 
The resulting increase of kinetic energy after this cycle  modifies the 
$\upmu$p atom diffusion in the hydrogen gas and the arrival time of the $\upmu$p atoms at the target walls.
This laser-induced modification of the arrival times is used to expose the atomic transition.
In this paper we present the simulation of the $\upmu$p diffusion in the H$_2$ gas which is at the core of the experimental scheme.
These simulations have been implemented with the Geant4 framework by introducing various low-energy processes including  the motion of the H$_2$ molecules, i.e. the effects related with the hydrogen target temperature.
The  simulations have been used to optimize the hydrogen target parameters (pressure, temperatures and thickness) and to  estimate  signal  and background rates.
These rates allow to estimate the maximum time needed to find the resonance and the statistical accuracy of the spectroscopy experiment. 
}

\vspace{10pt}
\noindent\rule{\textwidth}{1pt}
\tableofcontents\thispagestyle{fancy}
\noindent\rule{\textwidth}{1pt}
\vspace{10pt}

\section{Introduction}
\label{intro}

Highly accurate measurements of atomic transitions in muonic atoms can be used
as precise probes of low-energy properties of nuclei.
While the measurement of the $2s-2p$ transitions in light muonic atoms were used for the determination of nuclear charge radii and polarizability contributions \cite{Pohl:2010zza, Antognini:1900ns, Pohl1:2016xoo, Krauth:2021foz}, the measurement of the ground state hyperfine splitting  (HFS) in muonic hydrogen ($\upmu$p), an atom formed by a negative muon and a proton, can be used to advance the current understanding of the magnetic structure of the proton~\cite{Antognini:2022xoo, Peset:2021iul, Peset:2016wjq, Peset:2014jxa, Hagelstein:2015egb, Hagelstein:2015lph, Carlson:2011af, Tomalak:2018uhr, Karshenboim:2014vea, Faustov:2017hfo, Antognini:2022xqf}.

Three collaborations are aiming for the measurement of the HFS in muonic hydrogen with precision on the ppm level by means of pulsed laser spectroscopy~\cite{Pizzolotto:2020fue,Amaro:2021goz, Sato:2014uza}.
Comparing the measured HFS with the corresponding  theoretical prediction ~\cite{Antognini:2022xoo, Peset:2016wjq, Dupays:2003zz, Pachucki:1996zza} 
will result in a precise determination of the two-photon exchange contribution which is to be compared with predictions from chiral perturbation theory~\cite{Peset:2016wjq, Hagelstein:2015lph, Hagelstein:2018bdi}, and data-driven approaches based on dispersion relations~\cite{Carlson:2011af, Tomalak:2018uhr, Hagelstein:2015egb, Faustov:2006ve}.

The aim of this study (which is part of a PhD thesis~\cite{Jonas-Thesis}) is to describe and simulate the diffusion of  $\upmu$p atoms in hydrogen (H$_2$) gas which is at the core of the HFS experiment of the CREMA collaboration. The corresponding measurement will take place at a muon beamline of the CHRISP facility~\cite{CHRISP_Website} at the Paul Scherrer Institute (PSI) in Switzerland.
This study allows us to estimate background and signal rates, to optimize the experimental design
and to define the minimal requirements for the laser system, optical cavity and target.
Moreover, it enables an estimation of the maximum time needed to find the resonance and the statistical accuracy which can be reached in the HFS measurement.

In Sec.~\ref{sec:experiment} we introduce the principle and the experimental setup of the CREMA HFS measurement,  highlighting the role of the diffusion processes, which we aim to quantify in this study.
In Sec.~\ref{sec:collisions} we introduce the collisional processes between the $\upmu$p atom and the H$_2$ molecules and their related cross sections and collision rates. 
 How these collisional processes are implemented using Geant4  taking into account also the thermal motion of the H$_2$ molecules is presented in  Sec.~\ref{sec:implementation}.
An important input for the diffusion simulations is the kinetic energy distribution of the $\upmu$p atoms after their formation. This is discussed in  Sec.~\ref{sec:formation}.
In Sec.~\ref{sec:simulations} we present the results obtained from the diffusion simulations while in 
Sec.~\ref{sec:SigBGRates} we apply these simulations to determine signal and background rates and to optimize the target parameters.
Estimates of the  maximum time needed to search for the resonance and of the accuracy of the spectroscopy measurement can be obtained from the simulated signal and background rates, which is discussed in Sec.~\ref{sec:resonance}.

\section{The experimental scheme}
\label{sec:experiment}
\begin{figure}[bt]
  \centering
          \includegraphics[width=0.75\textwidth]{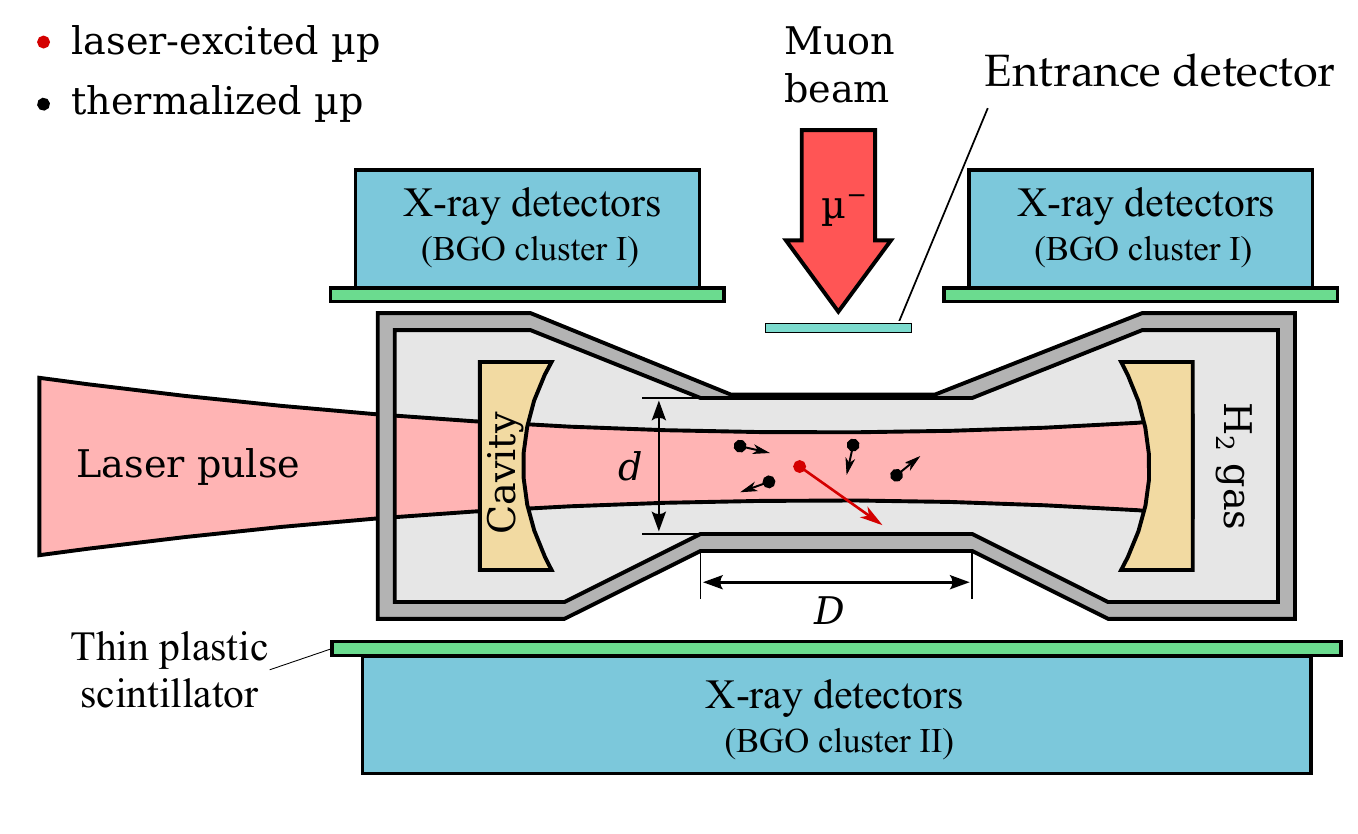}
\vspace{-4mm} \caption{Schematic (not to scale) of the setup for measuring the
  HFS transition. The $\upmu$p atoms are formed in a  hydrogen gas target (22~K, 0.6~bar) within a volume of thickness $d\approx 1$~mm (target thickness) and diameter $D\approx 15$~mm. If excited by the laser, the $\upmu$p atoms gain 0.1 eV kinetic energy and can diffuse efficiently towards the target walls. At the wall, the muons are transferred to gold atoms and form muonic gold atoms in highly excited states. The x rays produced during the subsequent deexcitation of the muonic gold atoms are detected outside of the hydrogen target using two clusters of Bismuth Germanium Oxide (BGO) scintillators.
  }
        \label{fig:target-region}
\end{figure}

The principle of the HFS experiment by the CREMA Collaboration is illustrated in  Fig.~\ref{fig:target-region}. 
A negative muon of about 11~MeV/c momentum passes an entrance detector triggering the laser system and is stopped in a cryogenic  H$_2$ gas target wherein $\upmu$p in a highly excited state is formed.
While the laser pulse is being generated, the $\upmu$p atom quickly deexcites to the 1$s$ state.
Through collisions with the H$_2$ molecules, the $\upmu$p atom ends up at the $F=0$ (singlet)  sublevel of the ground state while thermalizing to the H$_2$ gas temperature of 22~K. 
After $\sim 1~\upmu$s, the $\upmu$p atom is thermalized and the generated laser pulse of about 1~mJ energy at a wavelength of 6.8~$\upmu$m (corresponding to 0.18~eV) is coupled into a multi-pass toroidal cavity surrounding the muon stopping region.
The multiple reflections of the laser pulse  occurring in this cavity allow the illumination of a disk-shaped volume in the center of the target with a diameter of 15~mm and a thickness of 0.5~mm (see  Fig.~\ref{fig:target-region}). 
The on-resonance laser pulse may then excite the $\upmu$p atom from the singlet ($F = 0$) to the triplet ($F = 1$) sublevels. 
A successful excitation of the $\upmu$p atoms to the triplet state is followed within 10~ns by an inelastic collision between the $\upmu$p atom and a H$_2$ molecule  that deexcites  the $\upmu$p atom back to the singlet sublevel.
In this process, the HFS transition energy is converted into kinetic energy so that  the $\upmu$p atom acquires on average  0.1~eV of kinetic energy.
%
Because this energy is much higher than the thermal energy, the $\upmu$p atom  can quickly diffuse out of the laser-illuminated volume and reach  one of the target walls coated with gold.
When the $\upmu$p reaches the gold-coated walls, the negative muon of the $\upmu$p atom is transferred to a gold atom forming muonic gold ($\upmu$Au$^*$) in a highly excited state.
Through a cascade  of mainly radiative deexcitation, the $\upmu$Au$^*$ quickly reaches the ground state after the emission of several  x rays of MeV energy which can be detected as a signature of a successful laser excitation.
The HFS resonance can thus be obtained by counting the number of $\upmu$Au$^*$ cascade processes (referred in the following as $\upmu$Au events) in a certain time window (event time window) after the laser excitation as a function of the laser frequency.


\section{$\upmu$p-H$_2$ collisional processes}
\label{sec:collisions}

In this section we provide an introduction to the $\upmu$p$-$H$_2$ scattering processes relevant for this study. 
%
There are four molecular scattering processes relevant for  $\upmu$p atoms in the 1$s$ state which are classified according to the initial and final hyperfine states (total spin $F$), that can assume the values $F=0$ or $F=1$:
\begin{subequations}
  \label{eq:mol_scatt}
  \begin{alignat}{2}
    \label{eq:mol_scatt_11}
    & \upmu\mathrm{p}^{F=0}+\mathrm{H}_2 \rightarrow 
       \upmu\mathrm{p}^{F=0}+\mathrm{H}_2^{*} & \,, \\
    \label{eq:mol_scatt_12}
    & \upmu\mathrm{p}^{F=0}+\mathrm{H}_2 \rightarrow 
       \upmu\mathrm{p}^{F=1}+\mathrm{H}_2^{*} & \,, \\
    \label{eq:mol_scatt_21}
    & \upmu\mathrm{p}^{F=1}+\mathrm{H}_2 \rightarrow 
       \upmu\mathrm{p}^{F=0}+\mathrm{H}_2^{*} & \,, \\
    \label{eq:mol_scatt_22}
    & \upmu\mathrm{p}^{F=1}+\mathrm{H}_2 \rightarrow 
       \upmu\mathrm{p}^{F=1}+\mathrm{H}_2^{*} & \,.
  \end{alignat}
\end{subequations}
The superscript ``*'' indicates that  the rotational-vibrational state of H$_2$ can be altered by the scattering process.
Hence, none of these processes are strictly-speaking  elastic but we refer to the processes of Eqs.~(\ref{eq:mol_scatt_11}) and~(\ref{eq:mol_scatt_22}) as "elastic" in the sense that the total spin state $F$ of the $\upmu$p atom is conserved.

In the collisions described by Eqs.~(\ref{eq:mol_scatt}), the hyperfine state of the $\upmu$p atoms can be either conserved or changed by a spin-flip reaction,  in which the muon is transferred to a proton of the  H$_2$ molecule.  
These transfer reactions can thus lead to transitions between the two hyperfine levels depending on the spin of the proton to which the muon has been transferred. 
Note that the deexcitation rate of the upper spin state $(F=1)$  at the typical target conditions  is several orders of magnitude larger than the muon decay rate, which is in turn several orders of magnitude larger than the radiative deexcitation rate~\cite{Gershtein1958, Amaro:2021goz}.

Calculations of the differential cross sections for the processes in Eqs.~(\ref{eq:mol_scatt})(a-d) use the cross sections for the corresponding "nuclear" scattering processes of $\upmu$p on single protons, for which scattering amplitudes are available \cite{Bracci1989,Bracci1990,Bubak1987}. In addition, they take into account effects of molecular binding of the protons
in~H$_2$, electron screening and spin correlations for specific
rotational states of~H$_2$. A~method for calculating  the partial
differential cross sections of these processes is described
in~Refs.~\cite{Adamczak1996, Adamczak2006} and the numerical results for the cross sections are tabulated for      energies $\leq{}100$~eV
in Ref.~\cite{Adamczak1996}.
\begin{figure}[hbt]
  \centering
  \includegraphics[width=0.65\textwidth]{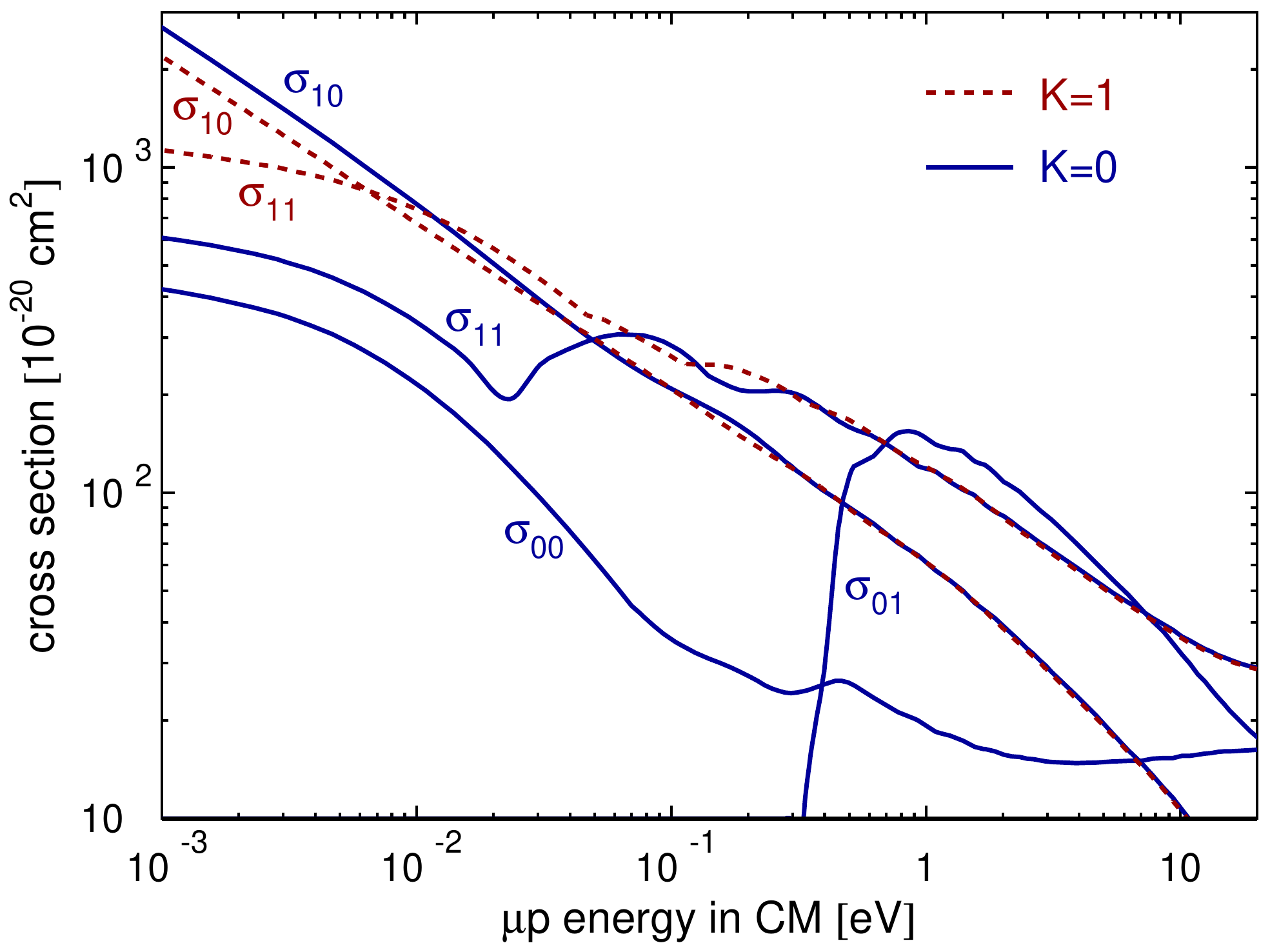}
  \caption{Total cross sections for the  processes given in Eqs.~(\ref{eq:mol_scatt}) as a function of the CM energy, for two
    initial rotational states of the hydrogen molecule: $K=0$ and $K=1$. The first  and second indices 
denote the initial and the final hyperfine states, respectively. Since the cross sections $\sigma_{00}$ and $\sigma_{01}$ for both initial rotational states  practically do not differ, these cross sections are plotted  for $K=0$ only.}
  \label{fig:xsections_cms}
\end{figure}
The total cross sections $\sigma^\text{CM}_\text{if}$  are given in
Fig.~\ref{fig:xsections_cms} as a function of collision energy in the center
of mass (CM) frame of the $\upmu\mathrm{p}+\mathrm{H}_2$ system. The indices $\text{i},\text{f}\in \{0,1\}$ denote the total spin of the initial and final hyperfine states, respectively. 
%
These cross sections account for all possible ro-vibrational excitations of the H$_2$ molecule.

The blue curves have been calculated assuming that all the H$_2$ molecules are
in the rotational state $K=0$ prior to a collision. 
For comparison, the cross sections $\sigma^\text{CM}_{10}$ and $\sigma^\text{CM}_{11}$ are also given assuming all
H$_2$ molecules are in the $K=1$ rotational state  prior to collision. 
As can be seen from the figure, minor differences between the two cross sections can only be found below
0.05~eV, which are due to the spin
correlations and different values of rotational thresholds for the two
initial rotational levels. It is also interesting to note that the cross section $\sigma_{01}$ has a threshold at a CM energy of $\sim 0.3$~eV corresponding to the minimum energy needed for upscattering the $\upmu$p atom from the lower lying singlet state to the triplet state.


\section{Monte Carlo simulations of molecular collisions}
\label{sec:implementation}

To simulate the diffusion process of $\upmu$p atoms in the H$_2$ gas, we have implemented the molecular scattering processes of Eqs.~(\ref{eq:mol_scatt})  as custom physics processes in the Monte Carlo toolkit Geant4~\cite{Geant4}, which we used via the program G4beamline~\cite{G4beamline}. 
In this section, we discuss how these physics processes were implemented using energy-dependent double-differential cross sections of the molecular collisions and taking into account the thermal motion of the H$_2$ molecules.

\subsection{Scattering rates in the laboratory reference system}
\label{sec:implementation:subsec:labRates}

To use the scattering cross sections discussed in Sec.~\ref{sec:collisions}, the cross sections need to be transformed from the center-of-mass to the laboratory (LAB) reference system. 
In this transformation,   the  thermal motion of the H$_2$ molecule can not be neglected as the $\upmu$p energy is similar to the thermal energy of the H$_2$ molecules.  
As the Geant4 program does not  account for the  motion (prior to collision)  of target particles, we define effective collision rates (from which the mean free path length needed by Geant4 can be deduced) that account for  the thermal motion of the H$_2$ molecules.

In the CM system, the total collision rates $\Gamma^\text{CM}_\text{if}$ are
defined as:
\begin{equation}
  \label{eq:rates_cms}
  \Gamma_\text{if}^\text{CM} = \rho v_\text{rel}\, \sigma^\text{CM}_\text{if} \,,
\end{equation}
where $\rho$ is the number density of H$_2$ molecules and $v_\text{rel}$
denotes the relative velocity of $\upmu$p and~H$_2$. 
The partial differential
rates $\partial^2{}\Gamma_\text{if}/(\partial{}E'\partial{}\Omega)$ in the LAB system
can be obtained from the calculated differential cross sections
$\partial^2{}\sigma_\text{if}^\text{CM}/(\partial{}E'_\text{CM}\partial{}\Omega_\text{CM})$ 
in the CM system:
\begin{equation}
  \label{eq:rates_lab}
  \frac{\partial^2{}\Gamma_\text{if}(E(v))}{\partial{}E'\partial{}\Omega} 
  = \frac{\rho}{2} \overline{ \sum_{E'_\text{CM}, \Omega_\text{CM}} 
    \int \mathrm{d}V \, g_M(V) \int_{-1}^{1} \mathrm{d}z_\alpha \,  
    v_\text{rel}(V,z_\alpha;v)\, \frac{\partial^2{}\sigma^\text{CM}_\text{if}(E_\text{CM})}
   {\partial{}E'_\text{CM}\partial{}\Omega_\text{CM}} 
   } \,\,,
\end{equation}
where $E$, $E'$, $\Omega$ and $E_\text{CM}$, $E'_\text{CM}$, $\Omega_\text{CM}$ denote
the initial and final $\upmu$p energies and the solid scattering angle in
LAB and CM system, respectively. The relative velocity, $v_\text{rel}$, depends on
the  $\upmu$p speed ($v$), on the  H$_2$ speed
($V$) -- both of them in the  LAB reference system -- and on $z_\alpha=\cos(\alpha)$, where $\alpha$ is
the impact angle. The  H$_2$ velocities are described by the
Maxwell-Boltzmann distribution $g_M$  at
a temperature~$T$.  The summation over $E'_\text{CM}$ and $\Omega_\text{CM}$
includes only contributions to the small intervals  $\delta{}E'$
around~$E'$ and $\delta{}\Omega$ around $\Omega$ from the differential
cross section in~CM system. Finally, the rate is averaged over a~distribution of the initial rotational energy levels for a~given H$_2$ target, which is denoted by the horizontal line.

\begin{figure}
  \centering
  \includegraphics[width=0.65\textwidth]{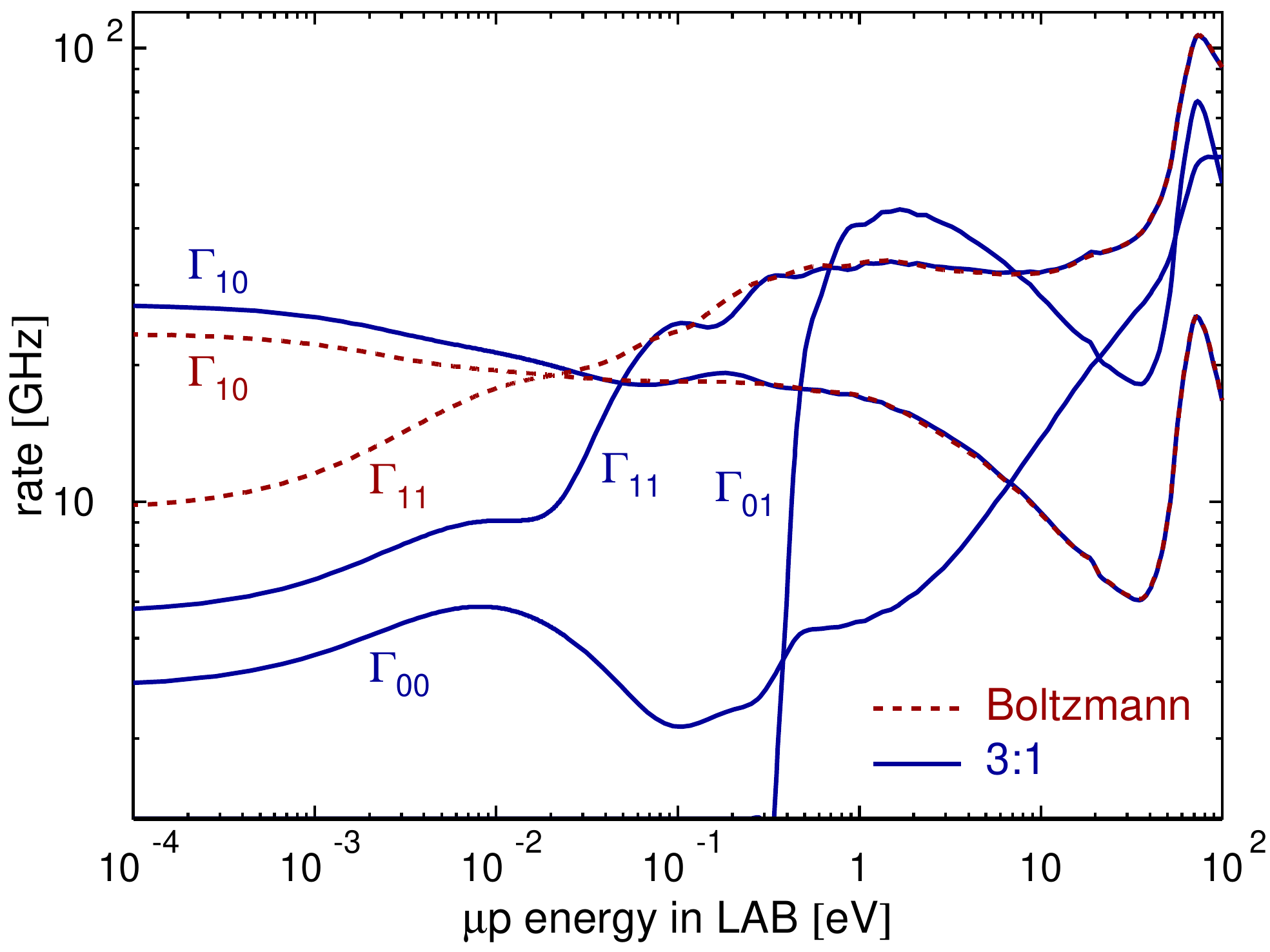}
  \caption{Collision rates $\Gamma_\text{if}$ for the processes given in Eqs.~(\ref{eq:mol_scatt}) calculated from Eq.~(\ref{eq:rates_lab})
    as a function of the $\upmu$p collision energy in the LAB reference frame at 22~K and liquid hydrogen density. The indices i and f denote the total spin of the initial and final hyperfine states, respectively. We considered both the Boltzmann and the 3:1 distribution of the initial rotational states of the H$_2$ molecules. Analogously to the cross sections $\sigma_{00}$ and $\sigma_{01}$ shown in Figure~\ref{fig:xsections_cms}, the rates $\Gamma_{00}$ and $\Gamma_{01}$ are not sensitive to the initial rotational distribution, and are therefore plotted  for the Boltzmann distribution only.
     }
  \label{fig:rates_lab}
\end{figure}

The calculated total rates~$\Gamma_\text{if}$ of the
processes given in Eqs.~(\ref{eq:mol_scatt})  are shown in
Fig.~(\ref{fig:rates_lab}) as functions of the $\upmu$p kinetic energy for a
target temperature of~22~K and for a liquid hydrogen density (LHD) $\rho_0=2.125\times{}10^{22}$~molecules/cm$^3$.
%
The results are shown for two distributions  of the initial rotational states of~H$_2$ molecules.
The Boltzmann distribution corresponds to the distribution of rotational states in hydrogen gas in thermal equilibrium. At 22~K this means that practically all H$_2$ molecules are in the state $K=0$. The 3:1 distribution instead denotes a mixture of the states $K=1$ and $K=0$ that corresponds to the degeneracy of both levels (3:1), as it is present in H$_2$ gas at room temperature. When cooling down hydrogen gas to 22~K, the gas will remain with the 3:1 rotational distribution for a long time (weeks at the conditions of the HFS experiment) because the relaxation process from $K=1$  to $K=0$ is very slow~\cite{Silvera:1980RevMod}. 

\subsection{Implementation in Geant4 of the molecular collisions at low energy }
For efficient implementation of the the LAB rates of Eq.~\eqref{eq:rates_lab} into Geant4, we export them in data tables. For each initial $\upmu$p energy, $E$, a table storing the double-differential cross section for various final energies, $E'$, and  scattering angles, $\theta$, is generated. 
In each $\upmu$p-H$_2$ scattering process, $E'$ and $\theta$ are sampled randomly from these two-dimensional distributions at the given energy $E$, while the azimuthal angle can simply be sampled from a uniform distribution.

All four molecular scattering processes of Eqs.~\eqref{eq:mol_scatt} were implemented as discrete processes in Geant4 using the following procedure. 
Each process proposes a  step length by randomly sampling from a Poissonian distribution with an expectation value given by the mean free path length.
The scattering process and the actual step length (distance travelled between two subsequent collisions) are then given by  the smallest of the proposed step lengths.

To calculate the mean free paths $\lambda_\text{if}$ for these processes at a given initial energy, $E$, we first calculate  the total rates $\Gamma_\text{if}(E)$ by integrating the right side of Eq.~(\ref{fig:rates_lab}) and then
use
\begin{equation}
  \label{eq:mfpFromRate}
\lambda_\text{if}\left(E\right) = \frac{v\left(E\right)}{\Gamma_\text{if}\left(E\right)},
\end{equation}
where $v(E)$ is the speed of the $\upmu$p atom in the LAB system having an energy $E$. 
%
%


\section{Kinetic energy of the $\upmu$p atoms after the cascade process}
\label{sec:formation}

For the simulations of $\upmu$p diffusion in the H$_2$ gas, knowledge of the $\upmu$p kinetic energy  right after the formation and deexcitation to the ground state is needed. 
However, the atomic formation process  and the subsequent deexcitation mechanisms are highly complex, which makes it difficult to assess the energy of $\upmu$p after reaching the 1$s$ state.

The $\upmu$p atoms are created in collisions of slow ($\sim 10$~eV) negative muons with H$_2$ molecules leading to the formation of $\upmu$p atoms in highly excited states with principal quantum number $n\approx14$~\cite{Fesenko1996}. 
The formation of the excited $\upmu$p atoms is followed by a fast ($\sim{}10^{-10}$~s for a density of $\varphi=10^{-2}$, given in fractions of the liquid hydrogen density $\rho_0$) deexcitation process (cascade) to the 1$s$ state, which involves radiative transitions, external Auger effects, Stark mixing, Coulomb transitions, elastic and inelastic scattering, and dissociation of H$_2$ molecules.
The cascade model on which this work is based, which shows a good agreement with experimental data, is summarized in Ref.~\cite{Covita:2017gxr}.

Apart from the radiative deexcitations, all
processes occur in collisions with surrounding H$_2$ molecules. Thus,
the rates of these processes depend on the target density, as well as on the collision
energy. 
For these reasons the most recent cascade model~\cite{Popov:2017PRA,Popov:2021PRA} also traces the kinetic energy evolution in addition to the evolution of the atomic state.
Indeed, the various cascade processes may lead to acceleration or deceleration of the excited $\upmu$p atoms.

The main acceleration of the $\upmu$p atoms occurs through Coulomb deexcitations in which the transition energy from $n$ to $n'$ is converted into kinetic energy shared between the colliding partners (the  $\upmu$p atom and an H$_2$ molecule).
In the Coulomb   $7\rightarrow 6$, $6\rightarrow 5$, $5\rightarrow 4$, $4\rightarrow 3$, and $3\rightarrow 2$ transitions with $\Delta n=1$, the $\upmu$p atoms acquire 9, 15, 27, 58 and 166~eV kinetic energy, respectively.
Coulomb deexcitations with $\Delta n>1$ lead to larger kinetic energies but are suppressed compared to $\Delta n=1$.
The acceleration from Auger processes is smaller as the electron carries away the largest fraction of the transition energy,
while  the radiative transitions do not significantly affect the kinetic energy of the $\upmu$p atoms as  the recoil energies are only of the order of few meV.
On the other hand, Stark transitions, elastic and inelastic collisions with or without excitation of the H$_2$ molecules are the processes which decelerate the $\upmu$p atoms down to thermal energies.

For our diffusion simulations  we chose an initial (right after the deexcitation to the $1s$-state) $\upmu$p energy distribution  that roughly approximates the  distribution given in Fig.~4 of Ref.~\cite{Covita:2017gxr}.
For simplicity, we assume  that \SI{50}{\percent} of the $\upmu$p atoms have a kinetic energy of 1~eV,  and the other \SI{50}{\percent} are uniformly distributed  between 0 and 100~eV. In the following, this initial energy distribution will be referred to as the $50/50$ distribution.
Even though this seems a crude approximation, it is sufficient for our purposes as the results do not depend on the details of the distribution.
The upper limit of 100~eV was imposed by the fact that no cross sections are available above this energy.
As we shall see later, the optimal target density for the HFS experiment is $0.008\leq \varphi\leq 0.01$ (given in fractions of the liquid hydrogen density), which is approximately a factor of 2 smaller than the density used to simulate the kinetic energy distribution reported  in Fig.~4 of Ref.~\cite{Covita:2017gxr}. Hence, for the HFS experiment, we expect in general a shift of the initial kinetic energies to lower values compared to Fig.~4 of Ref.~\cite{Covita:2017gxr}.
The kinetic energy assumed in this study is thus a conservative estimate yielding  conservative predictions of the signal rate.
To investigate the sensitivity of the results to the choice of the initial energy distribution and to give a lower limit  of the signal rate we also simulate the diffusion process assuming that all the $\upmu$p atoms have an initial kinetic energy of 100~eV.


\section{Simulations of the $\upmu$p diffusion in the HFS experiment}
\label{sec:simulations}

This section presents the simulation of $\upmu$p atom diffusion in the H$_2$ gas accounting for the low-energy processes
described above.
For clarity, this section is divided in three parts. The first describes $\upmu$p  thermalization and diffusion  before laser excitation (see Sec.~\ref{sec:prior-laser}). 
The second section provides some information on the laser excitation probability and depicts the kinetic energy distribution of the $\upmu$p atoms after a successful cycle of laser excitation and subsequent collisional deexcitation (see Sec.~\ref{sec:laser-excitation}).
The third section provides details on the diffusion of $\upmu$p atoms after laser excitation until they reach one of the target walls (see Sec.~\ref{sec:post-laser}).


\subsection{Thermalization and $\upmu$p diffusion prior to the arrival of the laser pulse}
\label{sec:prior-laser}

A muon passing the entrance detector triggers the laser system and then enters through a thin foil into the H$_2$ gas target with a thickness of about 1~mm in beam direction.
Depending on the gas density, only about 10-\SI{20}{\percent} (see later for more details) of the muons are eventually stopped in the H$_2$ gas forming $\upmu$p.
In this subsection we describe the thermalization of the $\upmu$p atoms after they have deexcited to the ground state, how they are quenched to the singlet sublevel, and how many of them remain in the hydrogen gas while the laser system is building up the pulse. 

Because the target is very thin compared to the width of the stopping range, in these simulations we generate the $\upmu$p atoms homogeneously distributed over the target thickness. 
We assume a stopping volume with a diameter $D= 15$ mm  (see Fig.~\ref{fig:target-region}) corresponding approximately to the muon beam size. We note that the results, for symmetry reasons, are  insensitive to the exact transverse distribution.
We also assume an initial kinetic energy distribution corresponding to the $50/50$ distribution described in Sec.~\ref{sec:formation}, and an initial population of the $\upmu$p hyperfine sublevels following the  statistical distributions: \SI{75}{\percent} in the triplet state and  \SI{25}{\percent} in the singlet state. 

\begin{figure}
\centering
\begin{subfigure}{0.49\textwidth}
\centering
\includegraphics[width=0.99\linewidth]{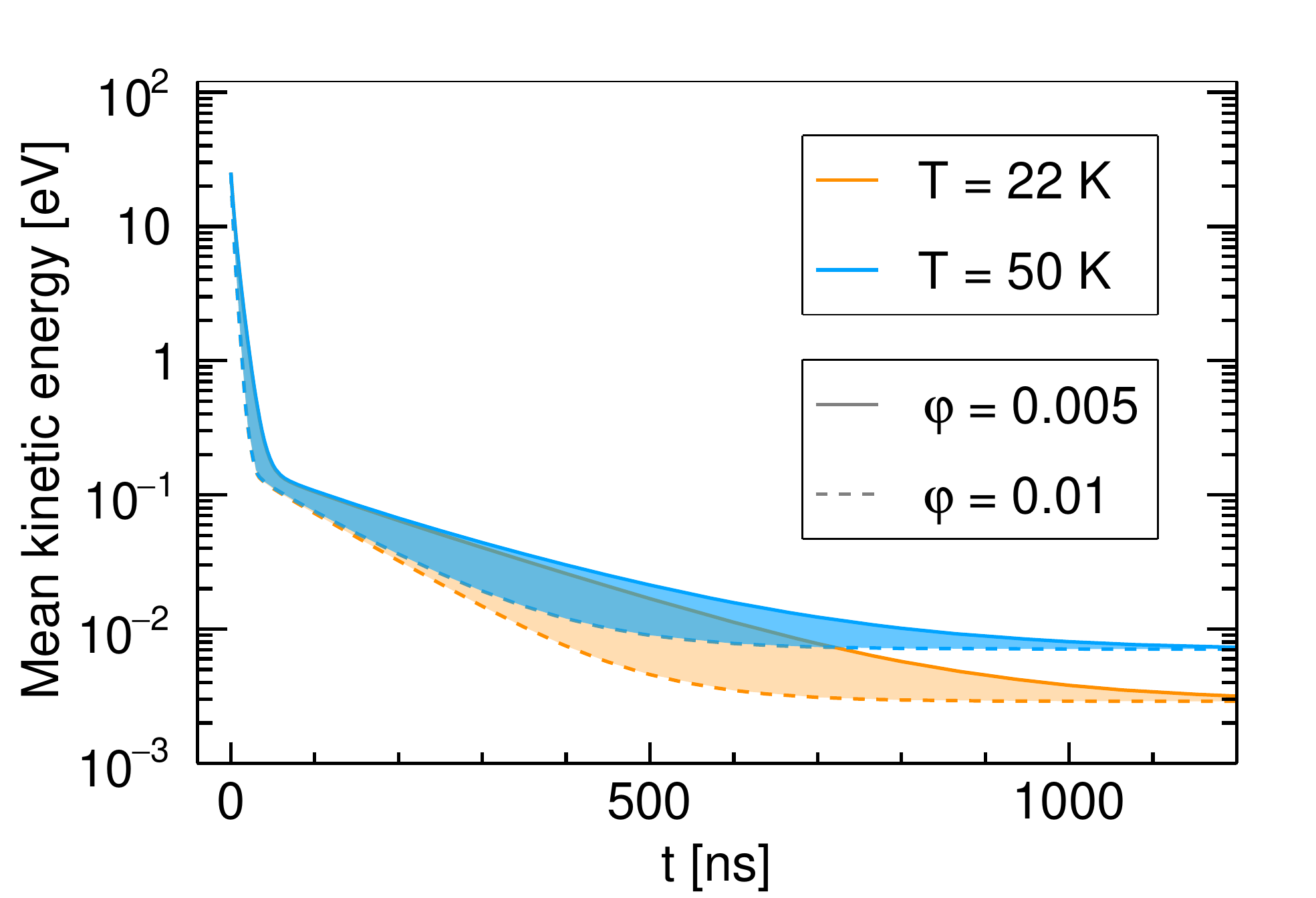}
\caption{}
\label{fig:lAS_a}
\end{subfigure}
\begin{subfigure}{0.49\textwidth}
\centering
\includegraphics[width=0.99\linewidth]{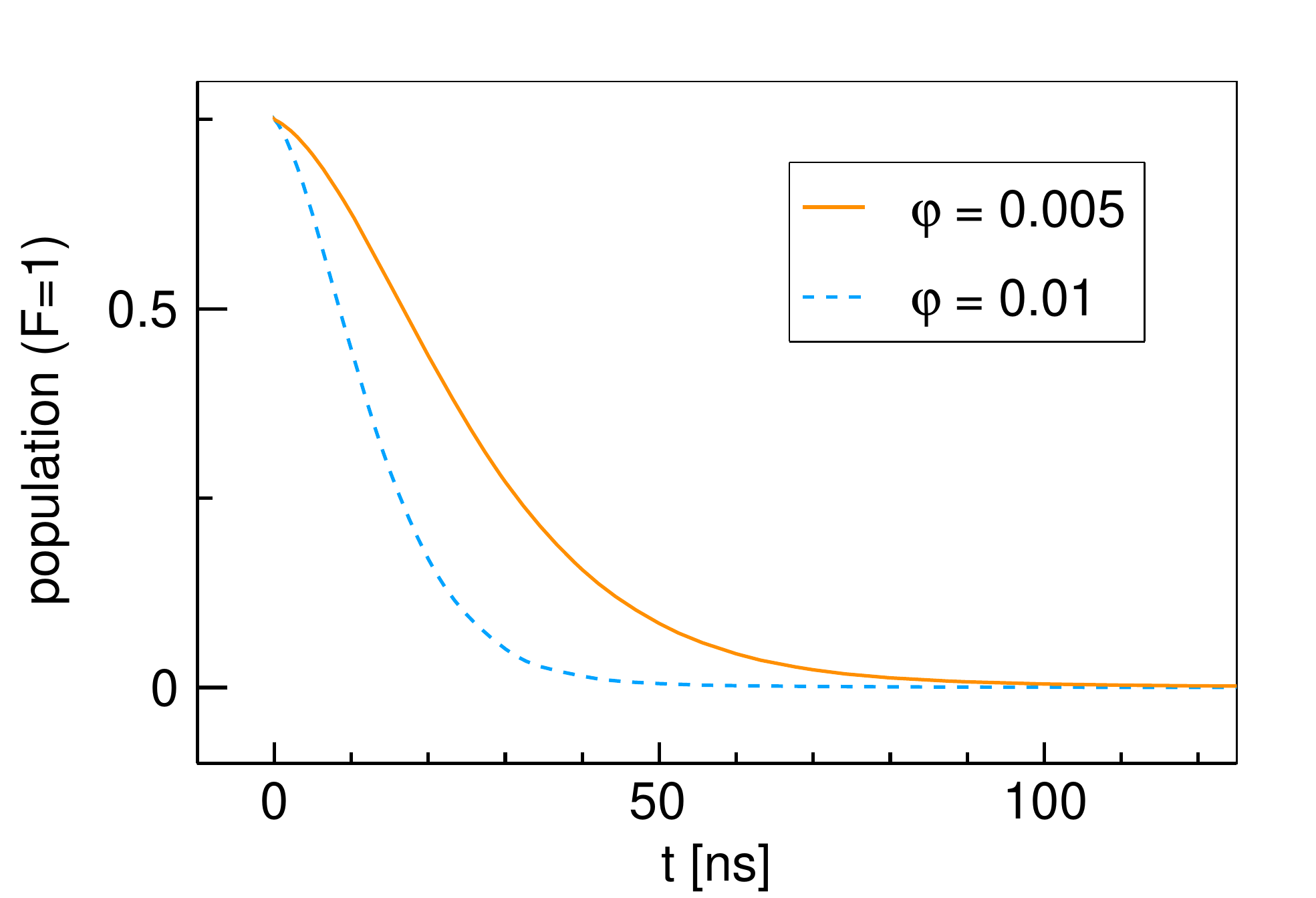}
\caption{}
\label{fig:lAS_b}
\end{subfigure}
\begin{subfigure}{0.49\textwidth}
\centering
\includegraphics[width=0.99\linewidth]{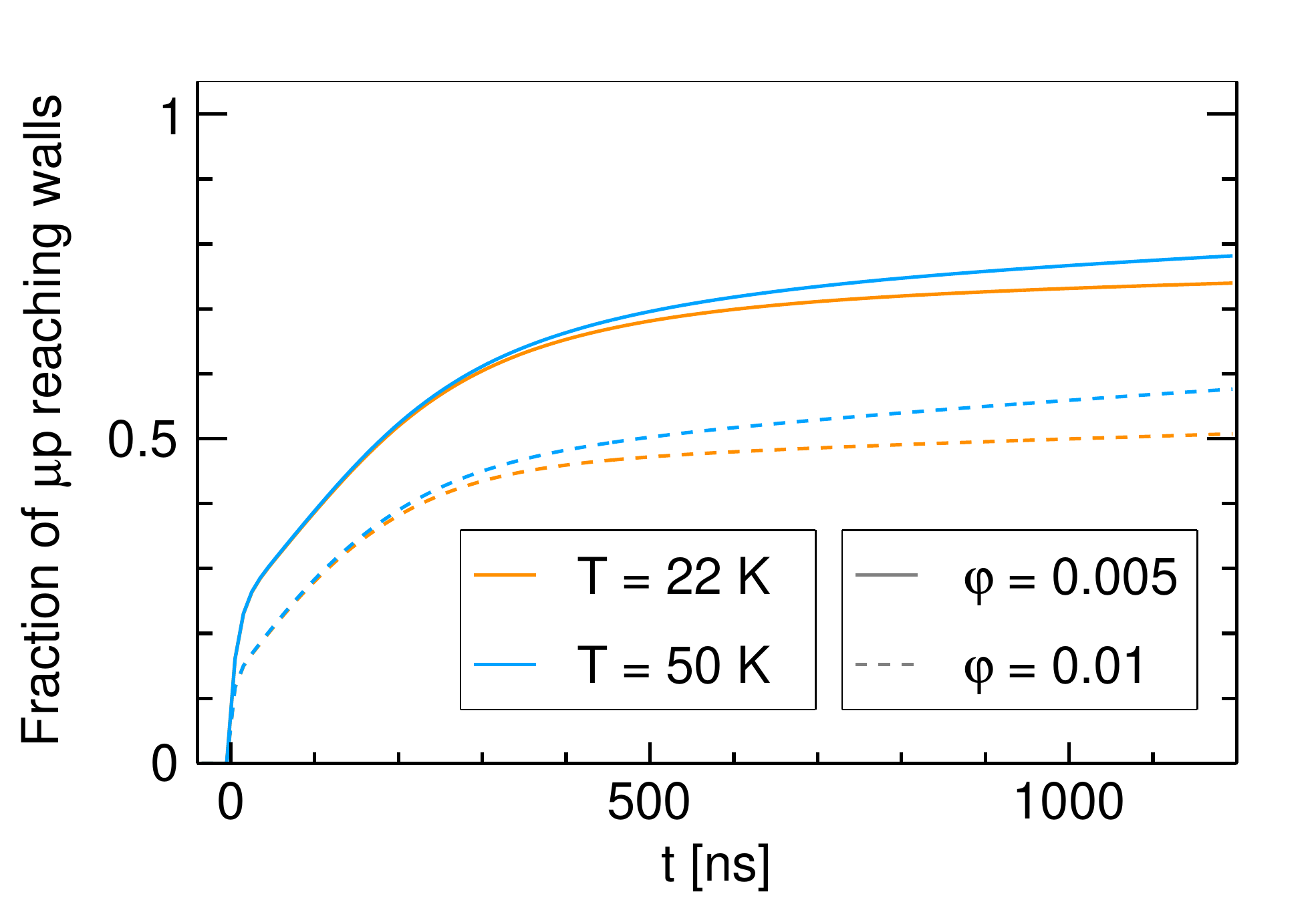}
\caption{}
\label{fig:lAS_c}
\end{subfigure}
\begin{subfigure}{0.49\textwidth}
\centering
\includegraphics[width=0.99\linewidth]{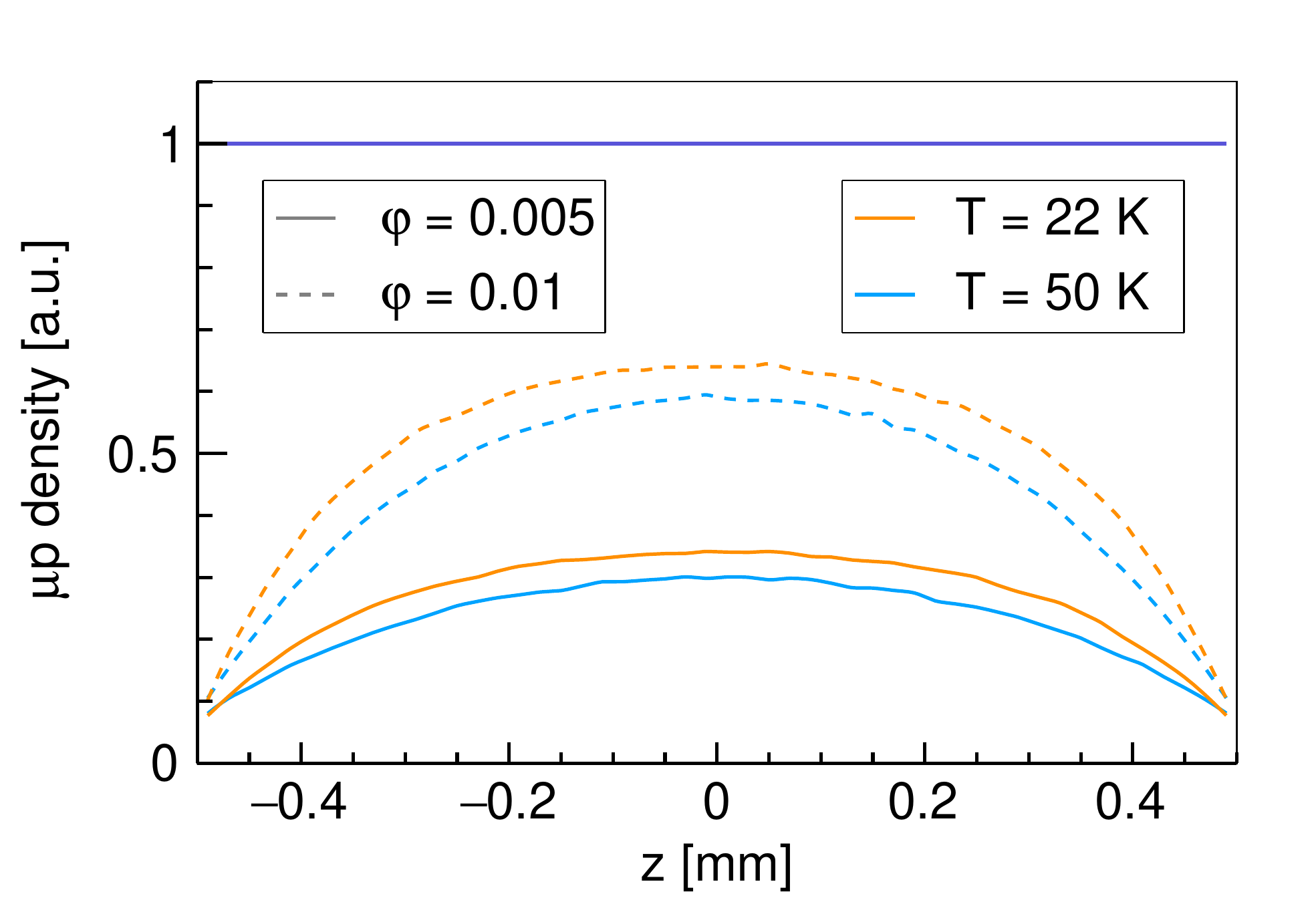}
\caption{}
\label{fig:lAS_d}
\end{subfigure}
\caption{(a) Time evolution of the average (LAB frame) kinetic energy of the $\upmu$p atom. (b) Time evolution of the population in the triplet  state. (c)  Fraction of $\upmu$p atoms that has reached either of the target walls. (d) Spatial distribution of the $\upmu$p atoms 1~$\upmu$s after their formation for a target of thickness $d=1$~mm. The horizontal line represents the distribution at time $t=0$ when the muonic atoms are formed. In all these plots we assumed that \SI{50}{\percent} of the $\upmu$p atoms have an initial kinetic energy of 1~eV and for the other \SI{50}{\percent} the initial energy is distributed uniformly between 0 and 100~eV. Initially, \SI{75}{\percent} of the $\upmu$p atoms are assumed to be in the triplet state.}
  \label{fig:lAS}
\end{figure}

Figure~\ref{fig:lAS_a} shows the evolution of the mean kinetic energy of the $\upmu$p atoms after the atomic cascade following the $\upmu$p formation,  simulated for various temperatures and densities $\varphi$. 
For example, a density of $\varphi=0.01$ corresponds to a pressure of about 0.6~bar at 22~K or 1.4~bar at 50~K. 
Throughout the paper we define as $t=0$ the time of the muonic atom formation. The cascade time is negligibly short on the $\upmu$s time scale of the experimental sequence.

As can be seen from the time evolution, within 1~$\upmu$s -- which corresponds to the time needed for the laser system to deliver the pulse -- the  $\upmu$p are thermalized at the H$_2$ gas temperature.
We anticipate here that the optimal target condition for a target of 1.0~mm thickness and 22~ K is $\varphi= 0.01$,
while  for a target  of 1.2~mm thickness it is $\varphi=0.008$.

The time evolution of the kinetic energy exhibits two slopes which are a consequence of the energy threshold of the up-scattering reaction from $F=0$ to $F=1$ described by $\Gamma_{01}$ (see Fig.~\ref{fig:rates_lab}). 
With mean energies far above the up-scattering threshold, both hyperfine sublevels are populated and there are continuous transitions between these two sublevels given by the rates $\Gamma_{01}$ and $\Gamma_{10}$. 
Hence, all four processes given in Eqs.~(\ref{eq:mol_scatt}) are occurring, leading to a fast energy loss and  a large slope in the kinetic energy plot.
When the energy drops below the up-scattering threshold of  $\sim 0.3$~eV, the up-scattering rate $\Gamma_{01}$ vanishes and all $\upmu$p atoms quickly deexcite to the $F=0$ state. 
A simulation of the $F=1$ population shown in Fig.~\ref{fig:lAS_b} confirms this behavior. Since the deexcitation happens at energies much higher than thermal energies, the given behaviour is not dependent on the temperature. 
The only collisional process that is still occurring  below this threshold is thus the  elastic scattering for the singlet sublevel with rate $\Gamma_{00}$, which is  the smallest compared to the other rates as visible from  Fig.~\ref{fig:rates_lab}.
For these reasons, for energies below the up-scattering energy threshold, the thermalization proceeds at a much slower pace.

At the target conditions of the HFS experiment, the $\upmu$p atoms with kinetic energies of a few tens of eV have mean free paths of a few hundred~$\upmu$m. 
Hence, a significant fraction of the $\upmu$p atoms formed in the hydrogen gas 
can diffuse into the target walls where the muon is transferred to the gold atoms and is  lost for the spectroscopy experiment. 
Figure~\ref{fig:lAS_c} shows the fraction of $\upmu$p atoms arriving at either of the target  walls versus time assuming a target thickness of 1~mm and that the initial energies of the $\upmu$p atoms follow the $50/50$ distribution discussed in Sec.~\ref{sec:formation}. 
As visible from the figure, between 50 and \SI{80}{\percent} of the total formed $\upmu$p atoms are lost prior to the laser excitation, depending on the target conditions. 
Note that temperature effects only emerge when the kinetic energy of the $\upmu$p approaches thermal energies. 
On top of this,  the losses due to muon decay (not included in these simulations) have to be considered: 
when the laser pulse arrives in the target at $t=1~\upmu$s, the number of surviving $\upmu$p atoms has been further reduced by a factor of $\exp{(-1/2.2)}=0.63$ (muon lifetime $\tau_{\upmu}=2.2~\upmu$s).

It is interesting to also consider how the spatial distribution of the $\upmu$p atoms evolves with time.
Figure~\ref{fig:lAS_d} shows the distribution of the $\upmu$p atoms along the 1~mm thick target at 1~$\upmu$s after the muonic atom formation for various density and temperature conditions.
The various curves  are normalized to the number of $\upmu$p atoms  at time $t=0$ (see blue horizontal line) when the muonic atoms are formed. Due to the low density of the hydrogen gas, the muon stopping distribution in the gas and the $\upmu$p distribution at $t=0$ were assumed to be flat.
Also in this plot, the decay losses (which do not modify the shape of the distribution) are not included and we assume the $50/50$ distribution for the initial energy of the $\upmu$p atoms.
The shape of the distribution changes with time (especially in first few hundreds of nanoseconds when the  $\upmu$p kinetic energies are large) because the walls of the target act as sinks for the $\upmu$p atoms, so that
the initial homogeneous distribution eventually develops a broad maximum
centered in the target mid-plane.

\subsection{Laser excitation}
\label{sec:laser-excitation}

After deexciting to the singlet state $F=0$ and reaching thermal equilibrium with the hydrogen gas, the $\upmu$p atoms are ready to be excited to the triplet state $F=1$ by the laser pulse. 
In Ref.~\cite{Amaro:2021goz} we calculated  the combined probability that a $\upmu$p atom -- initially in the singlet state and thermalized at H$_2$ temperature --  undergoes a laser transition to the triplet state followed by a collisional-induced deexcitation back to the singlet state.
This probability  accounts for de-coherence effects caused by  collisions and laser bandwidth as well as the Doppler broadening.

\begin{figure}
\centering
\begin{subfigure}{0.99\textwidth}
\centering 
\includegraphics[width=0.59\linewidth]{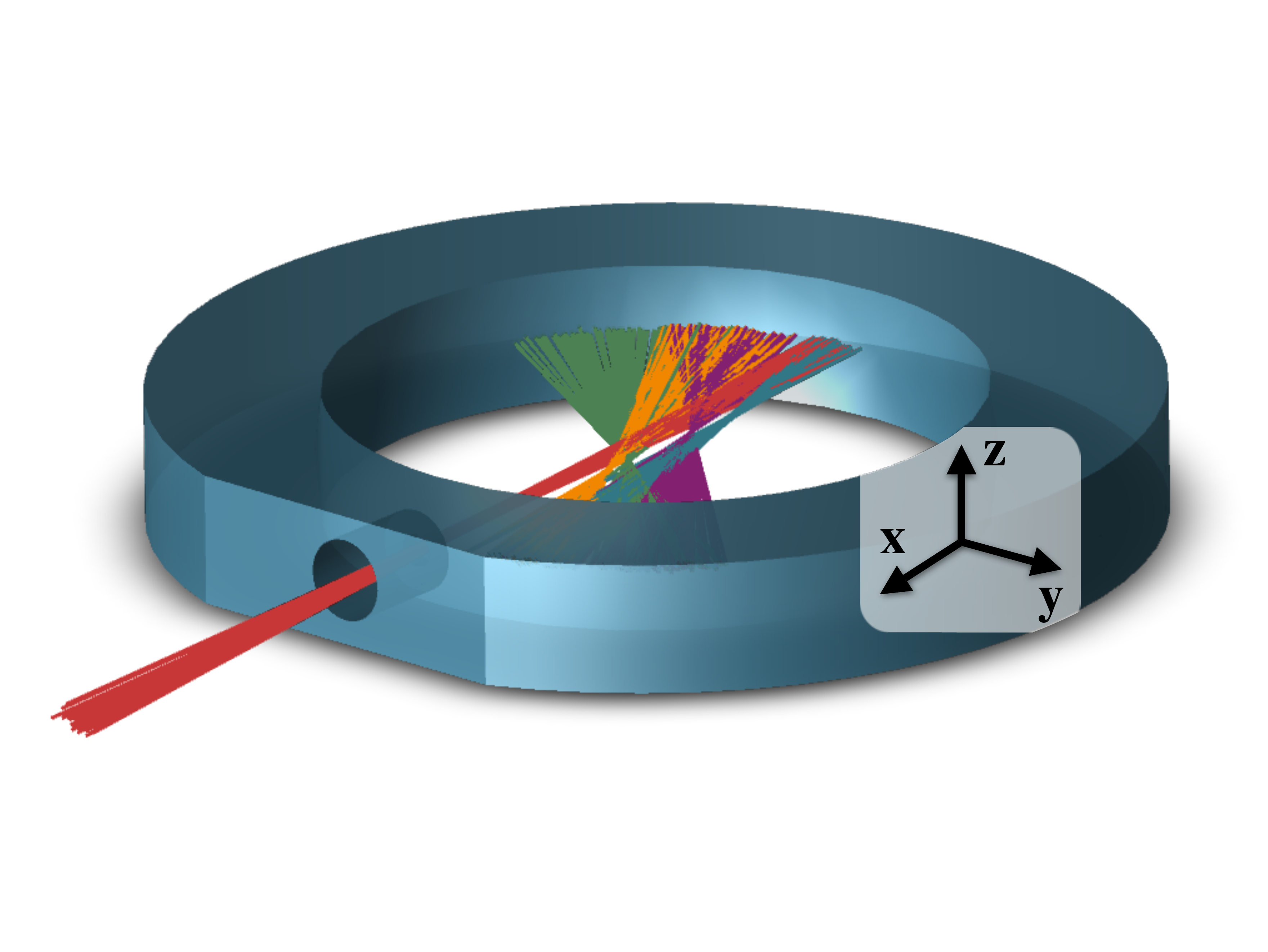}
\vspace{-13mm}
\caption{}
\label{fig:MirekCavity_a}
\end{subfigure}
\begin{subfigure}{0.49\textwidth}
\centering
\includegraphics[width=0.9\linewidth]{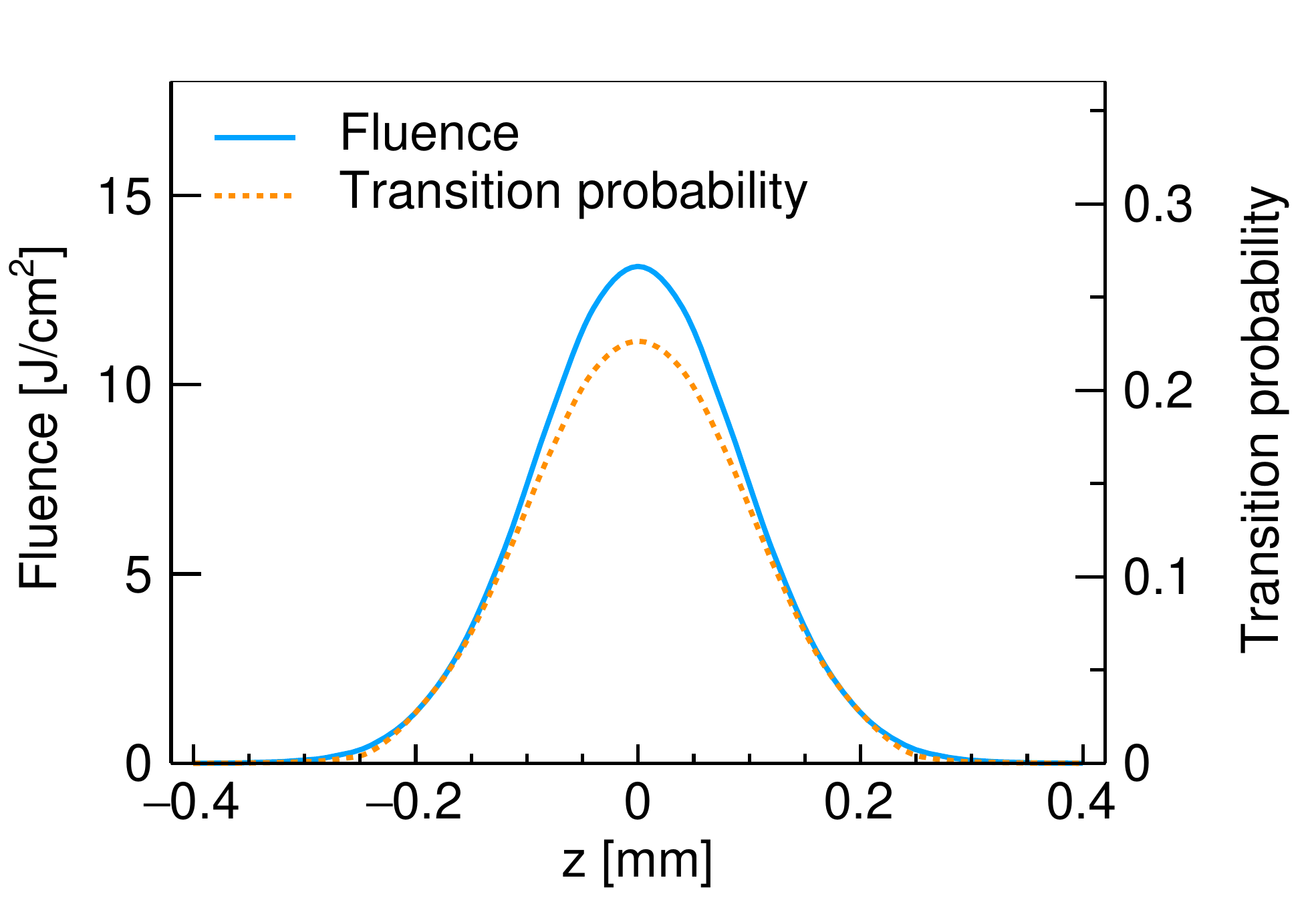}
\caption{}
\label{fig:MirekCavity_c}
\end{subfigure}
\begin{subfigure}{0.49\textwidth}
\centering
\includegraphics[width=0.9\linewidth]{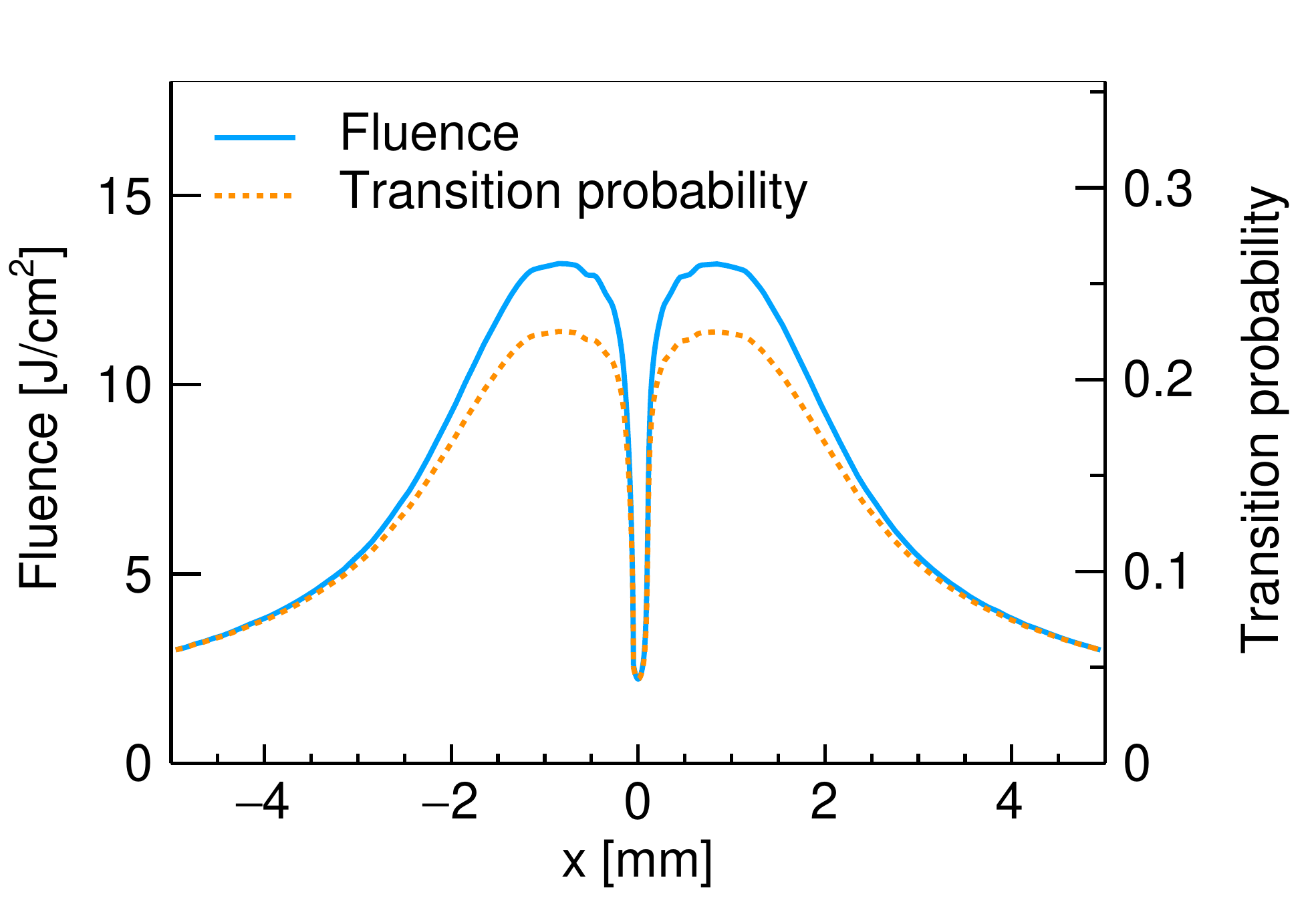}
\caption{}
\label{fig:MirekCavity_d}
\end{subfigure}
\caption{(a) 3D rendering of the toroidal multi-pass cavity with light rays. The laser light is injected into the cavity through a thin slit at a small angle relative to the x-axis. The light rays bounce back and forth within the cavity. To better illustrate the propagation of the ray bundels in the cavity, at each reflection the color of the rays is changed. (b) Spatial distribution of the fluence along the $z$-axis (direction of the muon beam). We assumed an in-coupled pulse energy of 1~mJ and a reflectivity of \SI{99.2}{\percent}. The plotted curve shows the fluence distribution passing through the maximum with a distance of $1~$mm from the target center, i.e. with $x=1~$mm. The laser-induced transition probability calculated from this fluence for a $\upmu$p atom in the $F=0$ state sitting in the respective position is also shown assuming additionally a laser bandwidth of 100~MHz, a pulse length of 50~ns, a target temperature of 22~K and a target pressure of 0.6~bar. Saturation effects are visible in the region of maximum fluence.  (c) Similar to (b) but in $x$-direction, i.e., in radial direction of the multi-pass cavity. }
  \label{fig:MirekCavity}
\end{figure}

To enhance the transition probability and to efficiently illuminate the muon stopping volume, the laser pulses are coupled into a multi-pass cavity.
We designed a toroidal multi-pass cavity, as shown in Fig.~\ref{fig:MirekCavity_a}, that allows the illumination of a disk-shaped volume, in which the light is coupled through a tiny slit and undergoes multiple reflections at the mirror surface.
The illuminated volume is matched to the geometry of the hydrogen gas target which has been chosen to be thin (in the $z$-direction) to allow the laser excited $\upmu$p atoms  to reach one of the target walls with high probability.
In contrast, the region of interest in the transverse direction is given by the size of the muon beam with a diameter of about 15~mm.
More details about the multi-pass cavity will be published elsewhere; here we give only the main features relevant for this study. Detailed studies of the fluence distribution and the performance of the cavity are included in Ref.~\cite{Mirek-Thesis}. Information about the laser system can be found, e.g., in Ref.~\cite{Manu-Thesis}.

\begin{figure}[t]
  \centering
  \includegraphics[width=0.65\textwidth]{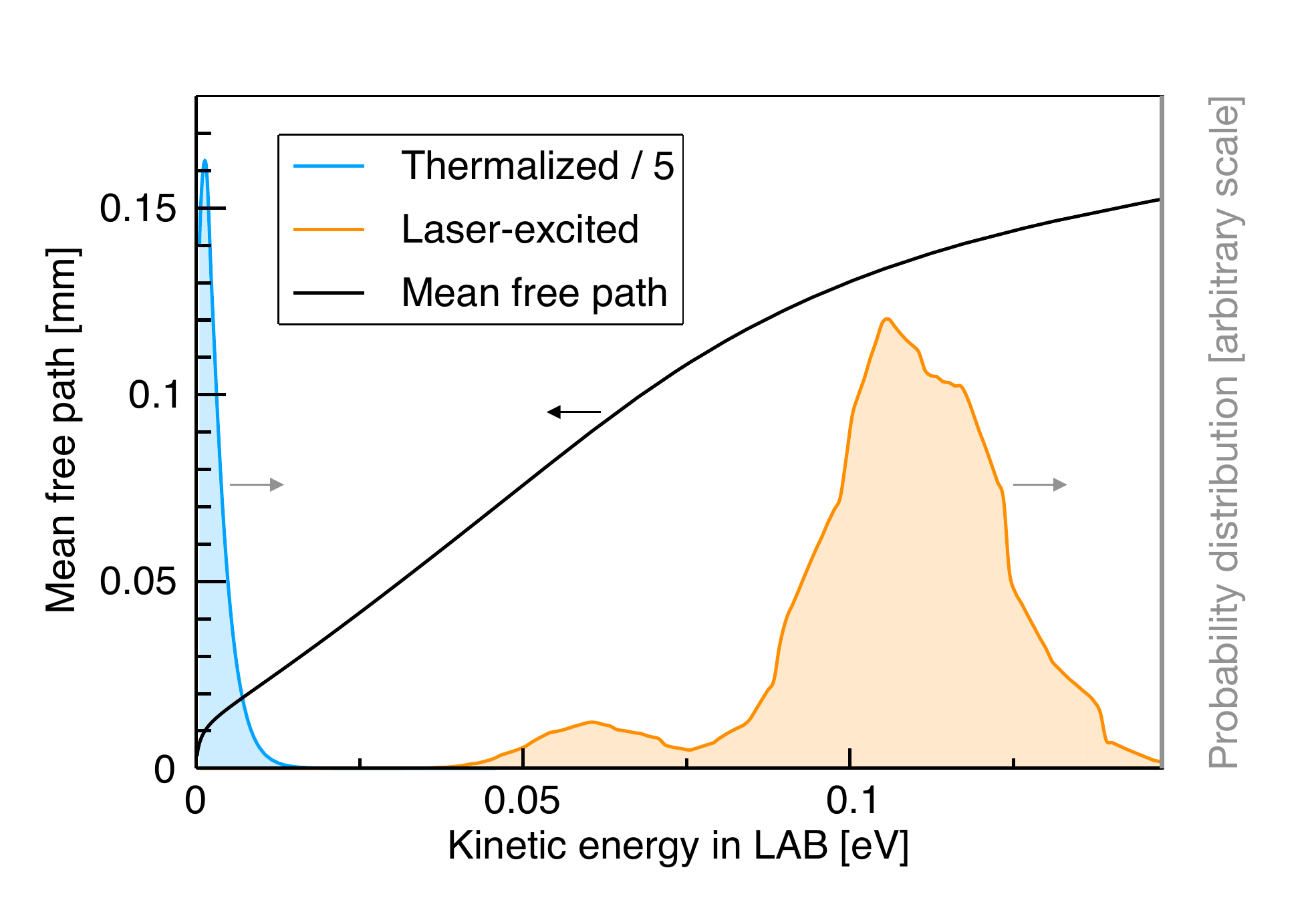}
  \caption{Kinetic energy distributions for thermalised $\upmu$p atoms (blue, rescaled in amplitude by a factor of 1/5) and for $\upmu$p atoms right after a successful cycle of laser excitation and collisional quenching from the triplet to the singlet state (orange). The mean free path versus kinetic energy of the $\upmu$p atoms is also given. Here we assume a target temperature of 22~K and a pressure of 0.6~bar.}
  \label{fig:spinflipDistribution}
\end{figure}

The fluence distribution of the toroidal cavity  has been simulated with a ray tracing program.
In the $z$-direction, the cavity is stable and the input beam is  mode-matched so that the light-distribution in this direction is Gaussian with a $1/e^2$ radius (waist)  of only 0.18~mm close to the center of the cavity (see Fig.~\ref{fig:MirekCavity_c}).
In the transverse plane the cavity is unstable so that the pulses bouncing back and forth in the cavity spread out with time,  resulting in a radial distribution as shown in Fig.~\ref{fig:MirekCavity_d}.
The fluence distribution  has been computed assuming 
an in-coupled pulse of 1~mJ energy, a mirror reflectivity of 0.992 (as measured for a copper cavity) and a cavity diameter of 100~mm.
At these conditions, the lifetime of the light in the  cavity is 40~ns to be compared with the laser pulse length expected to be around 30-50~ns.
The plots of  Figs.~\ref{fig:MirekCavity_c} and \ref{fig:MirekCavity_d} also depict the fluence-dependent excitation probability, showing small saturation effects even for relatively large fluences.
At the hydrogen gas conditions of the HFS measurement,  the laser-excited $\upmu$p atoms undergo an inelastic collision with a hydrogen molecule within 10~ns.
In this collisional induced deexciation, the $\upmu$p atoms get a share of the 0.18~eV transition energy, acquiring on average about 0.1~eV.
The exact kinetic energy distribution after the collisionally-induced deexcitation computed using the double-differential cross sections is shown in Fig.~\ref{fig:spinflipDistribution}.
For comparison, the kinetic energy distribution for thermalized $\upmu$p atoms (atoms which do not undergo the laser excitation) is also shown in the figure illustrating
 a clear separation in terms of kinetic energy for the two classes of $\upmu$p atoms: the ones which simply thermalize, and the ones which undergo a laser excitation.

Because the cross section $\sigma_{00}$ decreases with increasing energy, the additional energy kick won by the $\upmu$p atoms after a successful laser excitation allows them to efficiently diffuse in the H$_2$ gas.
To illustrate this, in Fig.~\ref{fig:spinflipDistribution}  we also plot the mean free path for $\upmu$p atoms in the H$_2$ gas as a function of the $\upmu$p kinetic energy as calculated from Eq.~\eqref{eq:mfpFromRate}.
Note that the minimum in the $\Gamma_{00}$ rates (see Fig.~\ref{fig:rates_lab}) coincides by chance approximately with the energy which is gained in the deexcitation process.

\subsection{Diffusion after the laser excitation}
\label{sec:post-laser}
The $\upmu$p atoms with extra kinetic energy from a successful laser excitation are generated in the central region of the target (around the target mid-plate at $z=0$). 
Their initial position distribution in $z$-direction is given by the excitation curve of Fig.~\ref{fig:MirekCavity_c} multiplied by the spatial distribution of the  $\upmu$p atoms at the moment of the laser excitation  given in Fig.~\ref{fig:lAS_d}.
Due to their relatively elevated kinetic energy after the collisional quenching and the small collision rate $\Gamma_{00}$, these $\upmu$p atoms travel comparably long distances and have a probability of reaching one of the target walls before thermalizing again of around \SI{30}{\percent} at optimal target conditions ($\varphi= 0.008-0.01$, $p= 0.5-0.6$~bar, $T=22$~K, $d=1.0$~mm).

As can be be read from the black curve of Fig.~\ref{fig:spinflipDistribution},
the $\upmu$p atoms with  the 0.1~eV kinetic energy can travel a distance of about 0.1~mm  before undergoing a  collision.
\begin{figure}[t]
\centering
\begin{subfigure}{0.49\textwidth}
\centering
\includegraphics[width=0.99\linewidth]{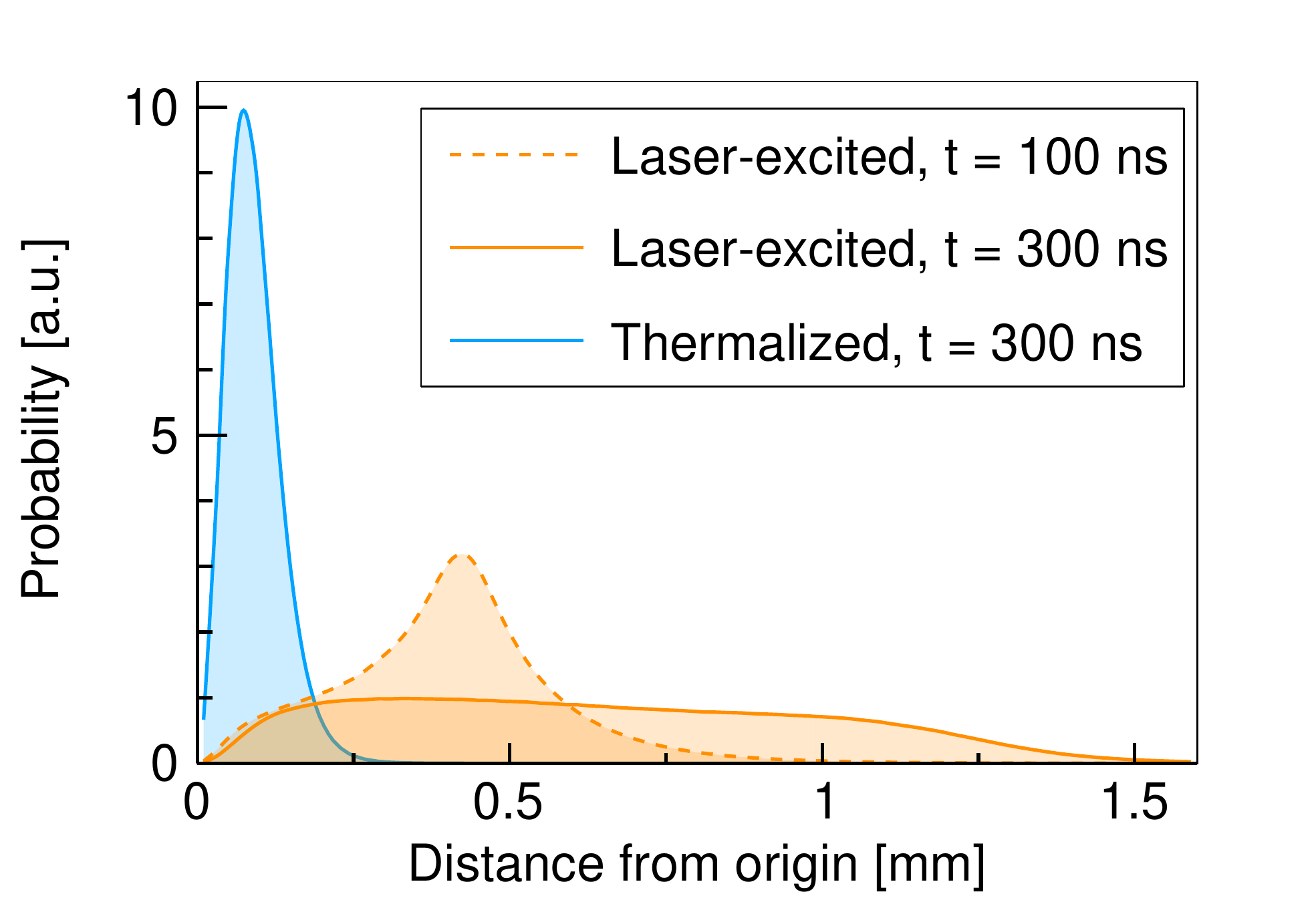}
\caption{}
\label{fig:diffusionRadius_a}
\end{subfigure}
\begin{subfigure}{0.49\textwidth}
\centering
\includegraphics[width=0.99\linewidth]{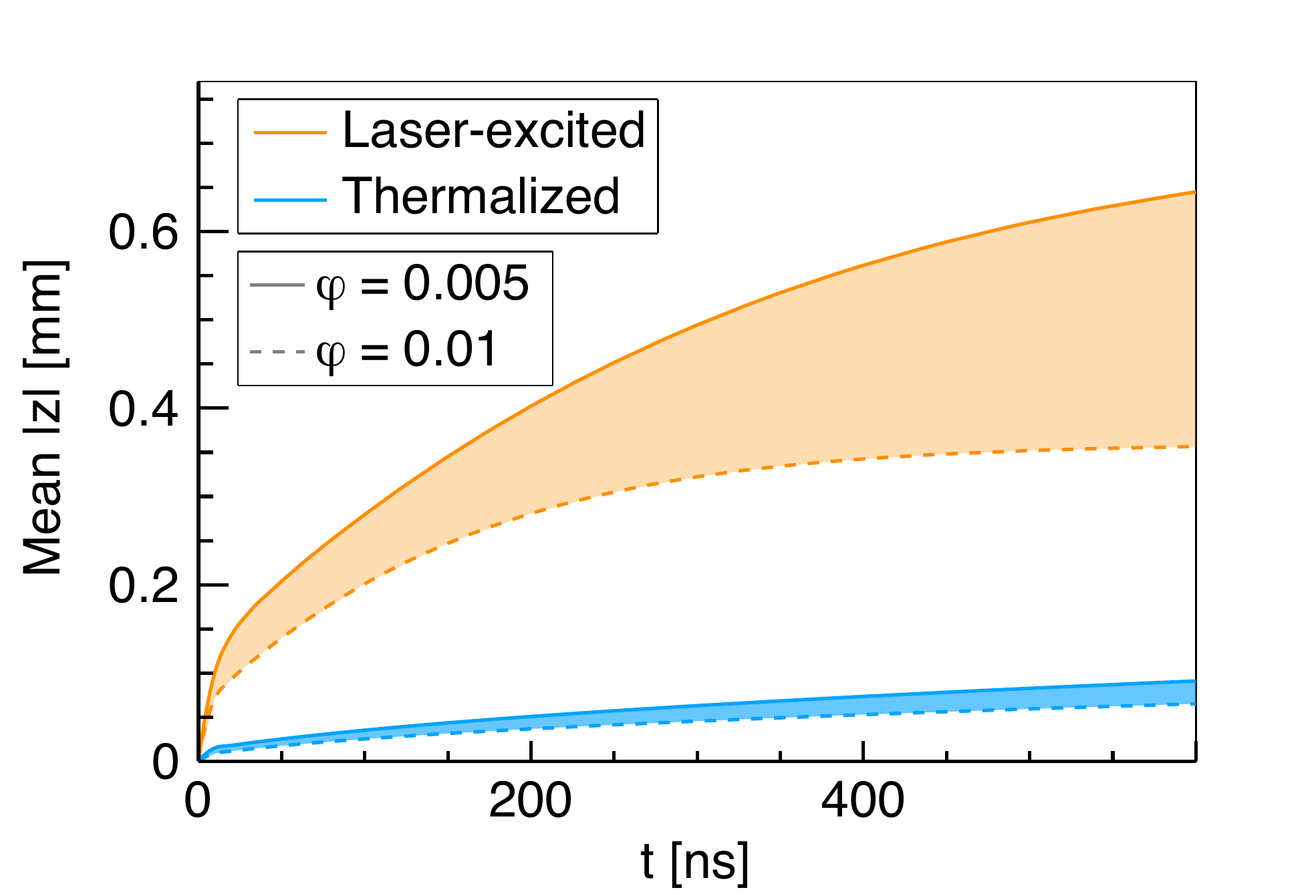}
\caption{}
\label{fig:diffusionRadius_b}
\end{subfigure}
\caption{(a) Distance travelled by the $\upmu$p atoms in the H$_2$ gas at  22~K temperature and 0.6~bar pressure in 100 and 300~ns, respectively. Two initial kinetic energies were assumed: for thermalized $\upmu$p atoms (blue) and for $\upmu$p atoms that have undergone a successful cycle of laser-excitation and collisional quenching (orange) as shown in Fig.~\ref{fig:spinflipDistribution}. (b) Mean distance in $z$-direction travelled by the $\upmu$p atoms as a function of time for two kinetic energy distributions and for two densities. }
  \label{fig:diffusionRadius}
\end{figure}
Yet, because the differential cross section is highest for small angles, a large fraction of these collisions lead to small energy losses and small angular deflections so that the $\upmu$p atom can efficiently diffuse for several hundred $\upmu$m before thermalizing.
Figure~\ref{fig:diffusionRadius_a} shows distributions of the distance travelled by $\upmu$p atoms in H$_2$ gas at 22~K and 0.6~bar ($\varphi=0.01$) at various times assuming an initial kinetic energy distribution as  obtained from the collisional deexcitation of the triplet state (see Fig.~\ref{fig:spinflipDistribution}).
For comparison, the distance travelled  by thermalized $\upmu$p atoms is also given.
As can be seen from this figure, within a few hundred nanoseconds, the $\upmu$p atoms that received the extra kinetic energy from the laser excitation travel  distances that are comparable to the  target thickness (ranging from 1 to 1.2~mm). 
On the contrary, the thermalized $\upmu$p atoms, i.e., $\upmu$p atoms that do not undergo laser excitation, travel maximally a few 100~$\upmu$m in the same time. 
Relevant for the experiment, the average distance the $\upmu$p atoms travel in the $z$-direction
is displayed in Fig.~\ref{fig:diffusionRadius_b}.
The  saturation visible  after a few hundred nanoseconds is because most of the $\upmu$p atoms have lost their excess energy.
%
\begin{figure}
\centering
\begin{subfigure}{0.49\textwidth}
\centering
\includegraphics[width=0.99\linewidth]{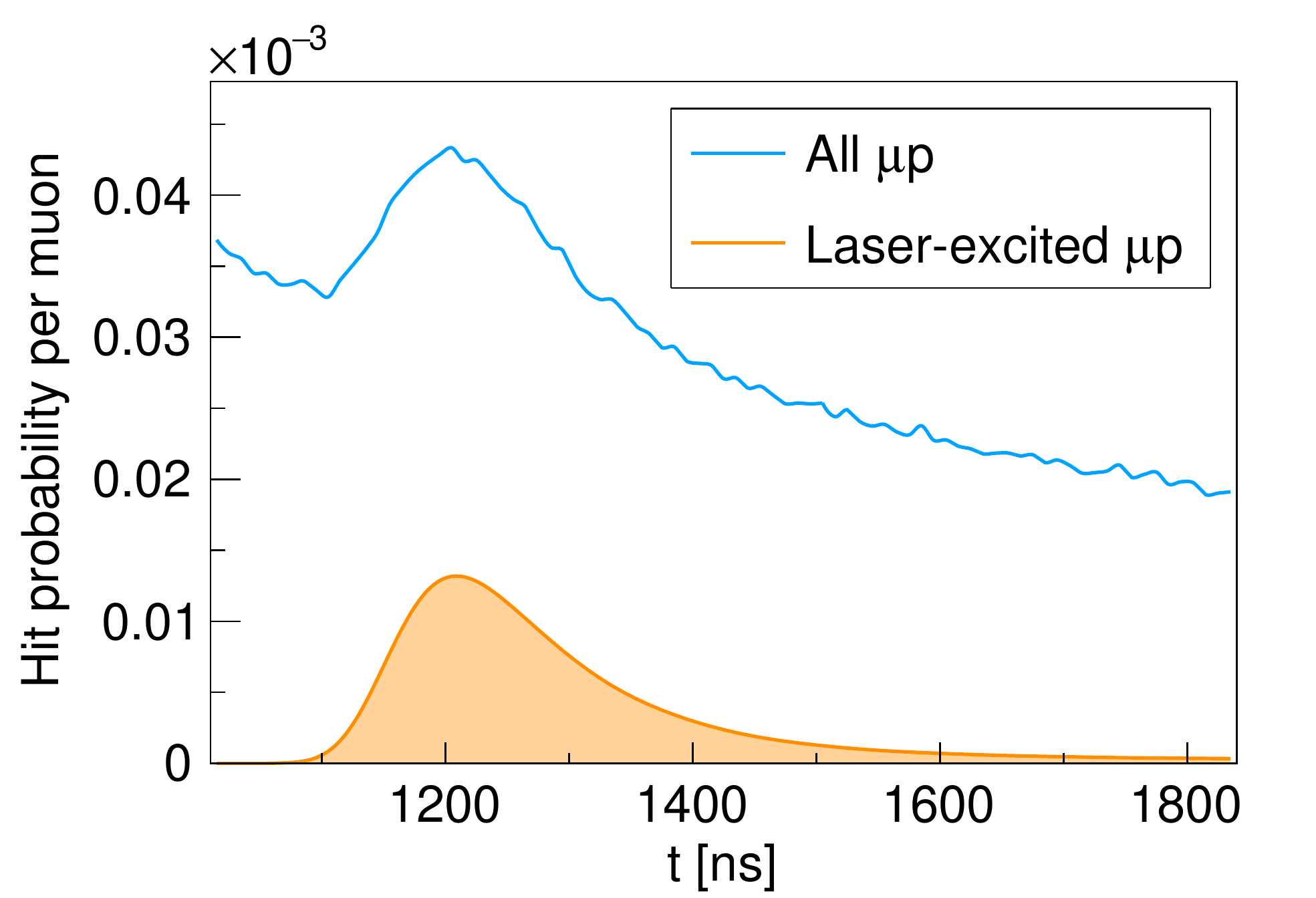}
\caption{}
\label{fig:finalHits_a}
\end{subfigure}
\begin{subfigure}{0.49\textwidth}
\centering
\includegraphics[width=0.99\linewidth]{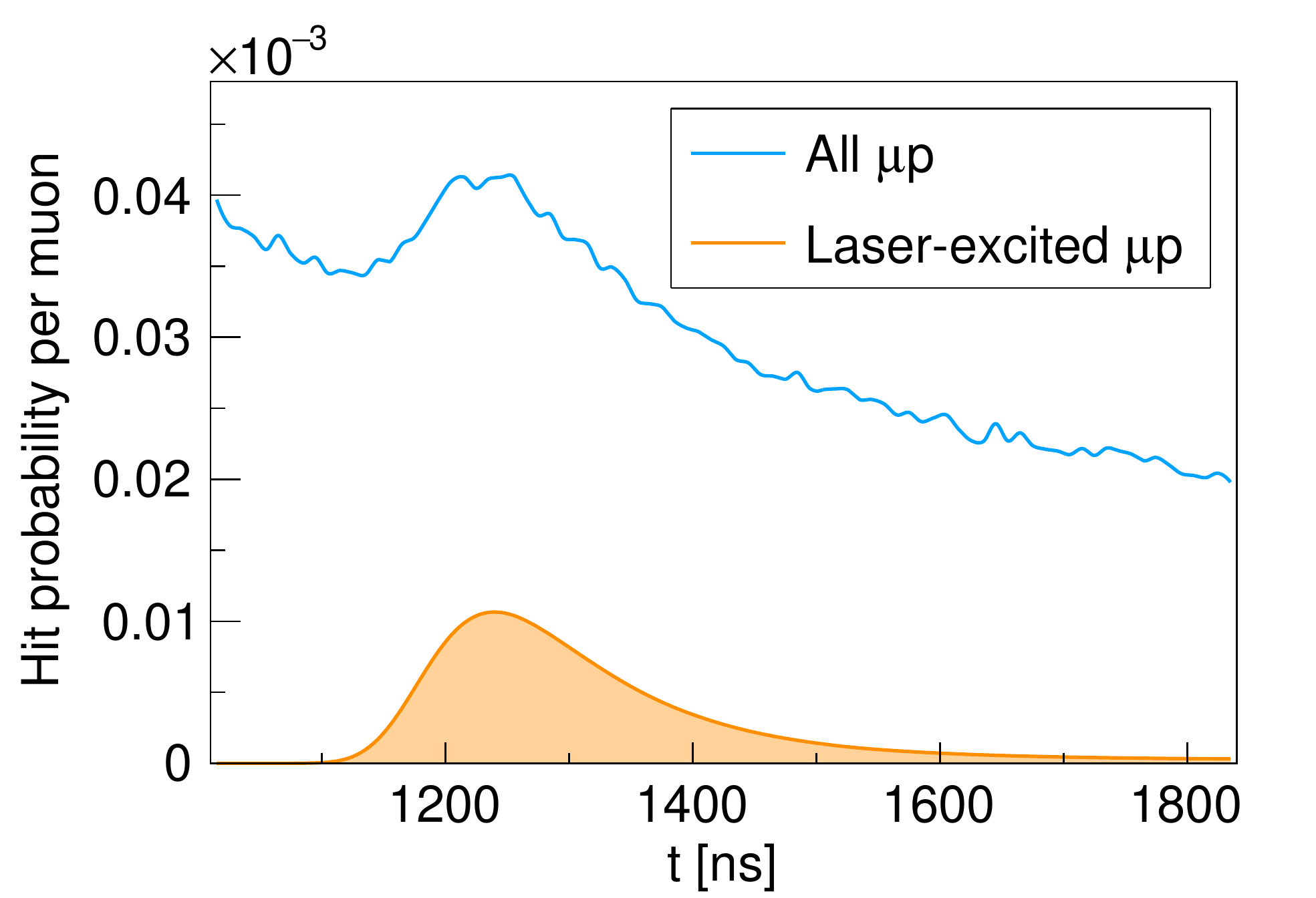}
\caption{}
\label{fig:finalHits_b}
\end{subfigure}
\begin{subfigure}{0.49\textwidth}
\centering
\includegraphics[width=0.99\linewidth]{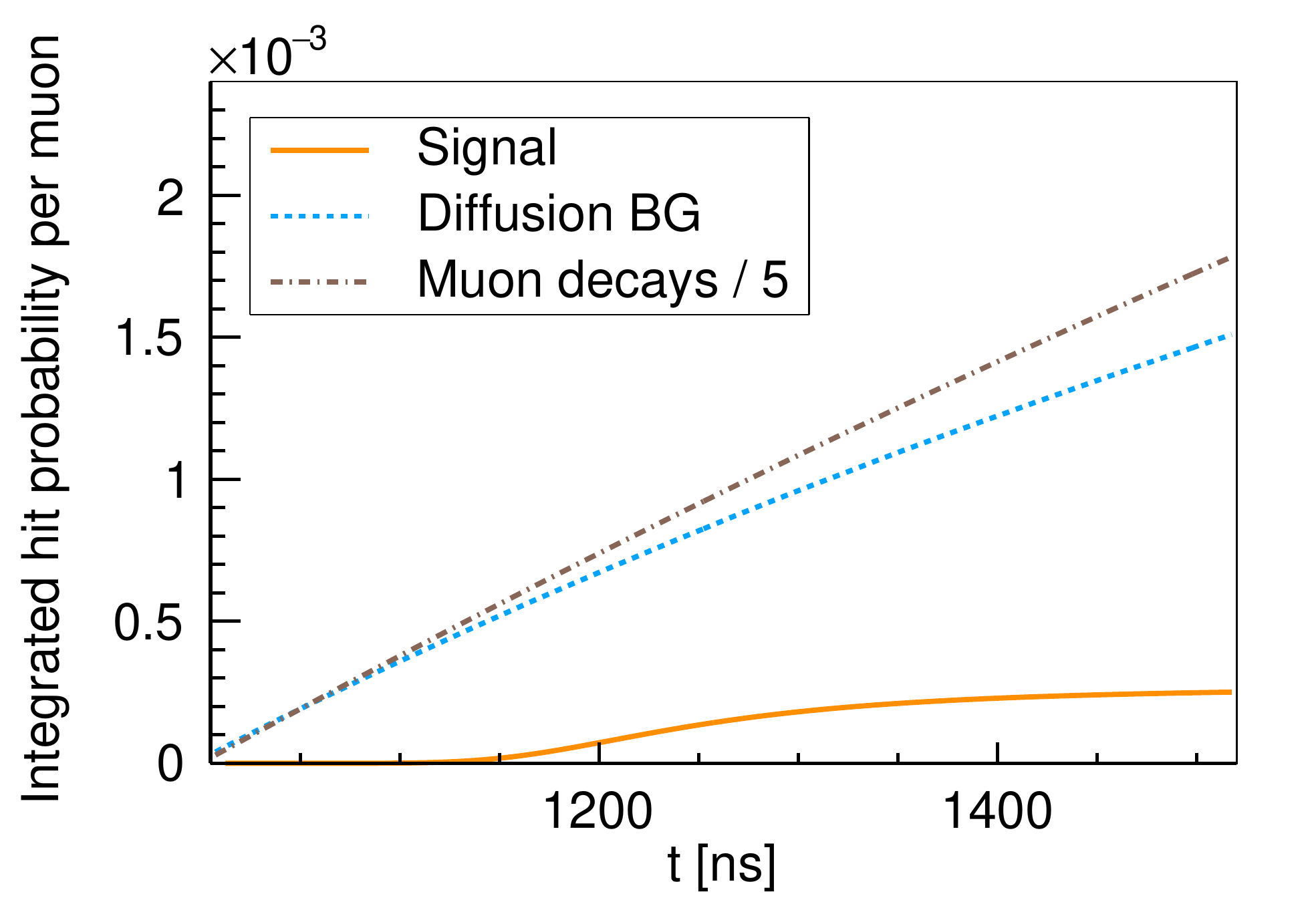}
\caption{}
\label{fig:finalHits_c}
\end{subfigure}
\begin{subfigure}{0.49\textwidth}
\centering
\includegraphics[width=0.99\linewidth]{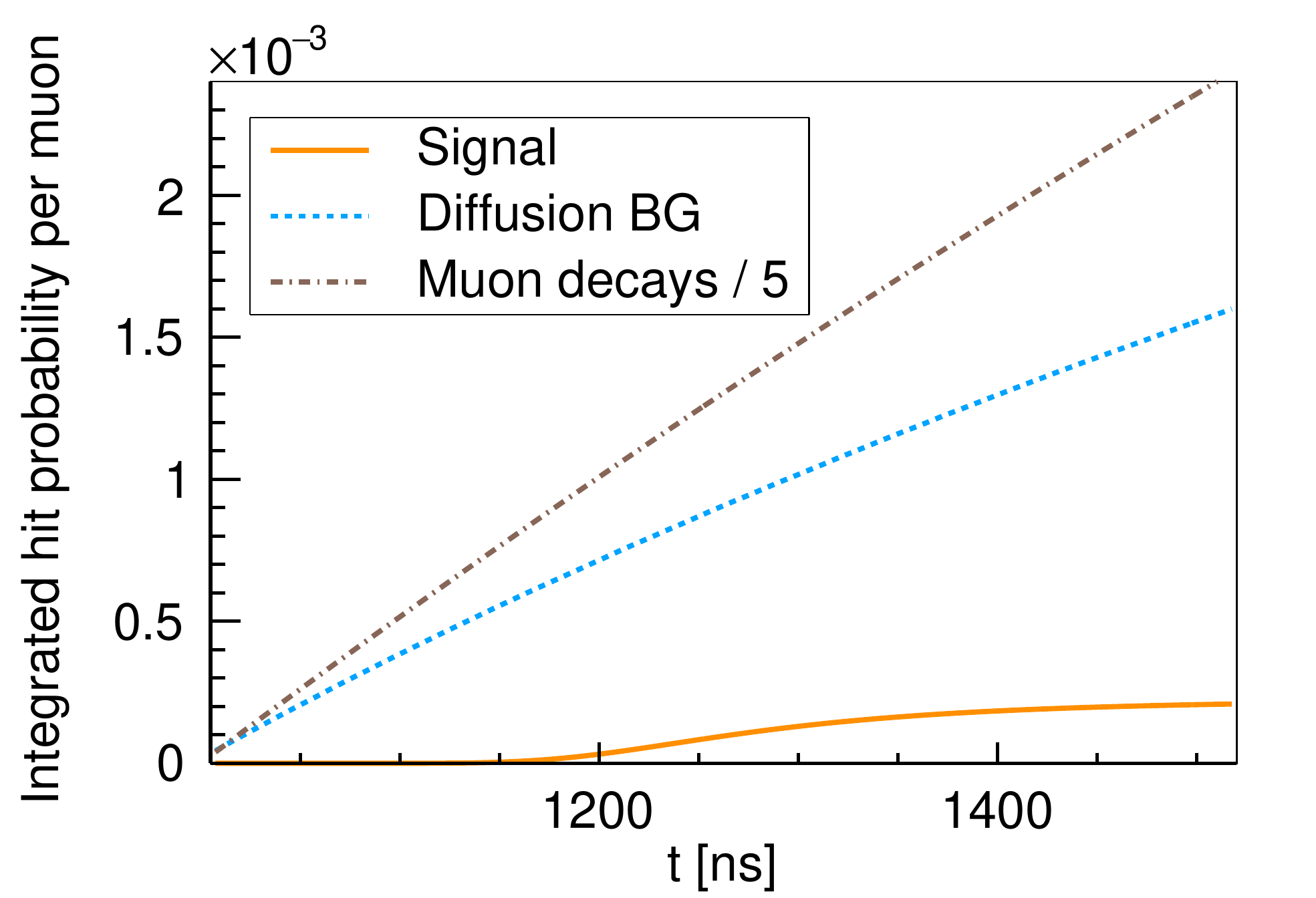}
\caption{}
\label{fig:finalHits_d}
\end{subfigure}
\caption{(a) Time distribution of the $\upmu$p atoms arriving at one of the target walls for times following the laser excitation. In orange we show the arrival times only for $\upmu$p atoms that have undergone laser excitation, in blue for all $\upmu$p atoms. We assume the $50/50$ initial $\upmu$p energy distribution described in Sec.~\ref{sec:formation} and a hydrogen gas target with 22~K temperature, 0.6~bar pressure and 1.0~mm thickness. We further assume that at $t=1000~$ns, laser pulses of 1~mJ energy, 50~ns pulse length and 100~MHz bandwidth are coupled into the multi-pass cavity having a reflectivity of \SI{99.2}{\percent}. 
The hit probabilities have been normalized to the number of muons entering the target and binned  in 10~ns.
(b) Similar to (a) but for a target thickness of 1.2~mm. (c) and (d) Integral from $1000$~ns to $t$ of the time spectra depicted in Figs.~\ref{fig:finalHits_a} and  \ref{fig:finalHits_b}, respectively. The dark brown (dashed-dotted) curves represent the probability per entering muon that a $\upmu$p atom is decaying in the time window between 1000~ns and $t$. Note that this probability has been divided by 5 to increase readability.
}
  \label{fig:finalHits}
\end{figure}

Figures~\ref{fig:finalHits_a} and~\ref{fig:finalHits_b} show the time distributions of $\upmu$p atoms reaching one of the walls, assuming that the laser pulses are coupled into the multi-pass cavity at time $t=1000$~ns and that the $\upmu$p atoms are formed at time $t=0$.
These simulations use the results from the diffusion simulation prior to the laser excitation and the results from  the laser excitation section.
In both cases we assumed the $50/50$ distribution for the initial $\upmu$p energy and we simulated a target with 0.6~bar pressure and 22~K temperature.
Only the target thickness differentiates the two simulations: for the left plot we used a thickness of $d=1.0$~mm, for the right plot we used a thickness of $d=1.2$~mm.
Each plot shows two curves: the orange curves represent the distribution of arrival times for $\upmu$p atoms that   have undergone the laser excitation, while the blue curves represent the arrival times for all $\upmu$p atoms independently of whether or not they have undergone  the laser excitation.
As visible from the orange curves,  it takes about 100~ns until the first $\upmu$p atoms which have undergone laser excitation reach one of the target walls.  Most of them however arrive at the walls in  a time window spanning from  about 1100 to 1500~ns (that depends on the target thickness).
This spread of arrival times originates from the initial kinetic energy distribution, the isotropic distribution of the initial velocity, the spatial distribution of the $\upmu$p atoms at the time of the laser excitation and the collisional effects during the diffusion process.

By considering the blue curves of Figs.~\ref{fig:finalHits_a} and~\ref{fig:finalHits_b}, we notice that there is a large fraction of $\upmu$p atoms arriving at the target walls without being excited by the laser pulse. 
These $\upmu$p atoms give rise to background events as each $\upmu$p atom arriving at the target walls leads to a $\upmu$Au event, independently of whether previously they had undergone or not a laser excitation.
This background is produced by $\upmu$p atoms that have thermalized in the vicinity of the target walls and that slowly diffuse to  one of the walls in the event time window within which laser-induced $\upmu$Au events are recorded.

The  simulations of Figs.~\ref{fig:finalHits_a} and~\ref{fig:finalHits_b} are normalized by the number of incident muons and  account for the stopping probability in the hydrogen gas, $P_{\text{stop}} $, which was estimated conservatively based on measurements using the relation
\begin{equation}
P_{\text{stop}} [\%]=500\cdot \frac{p\left[\text{bar}\right]\cdot d\left[\text{mm}\right]}{T \left[\text{K}\right]}\text{,}
\end{equation}
where $T$ is the temperature of the gas in K, $p$  the target pressure  in bar, and $d$ the  target thickness in mm.

The time window in which the $\upmu$Au events have to be searched for (event time window), has to be selected as a trade-off between signal and background rates.
Figures \ref{fig:finalHits_c} and  \ref{fig:finalHits_d} represent the integral from $1000$~ns to $t$ of the time spectra depicted in Figs.~\ref{fig:finalHits_a} and  \ref{fig:finalHits_b}, respectively.
Hence, the asymptotic value of the orange curves of Figs.~\ref{fig:finalHits_c} and ~\ref{fig:finalHits_d} represents the probability for a laser-induced $\upmu$Au event per entering muon.
As we will see in the next section, these integrals allow to quickly determine signal and background rates in a given event time window.
On top of the above described diffusion-related background visible in Figs.~\ref{fig:finalHits_a} and~\ref{fig:finalHits_b}, there are two other sources of background that we have to consider in the spectroscopy experiment. 
One is a muon-uncorrrelated background, i.e., a background originating from natural radioactivity in the experimental area, from neutrons produced by the accelerator facility, from possible electron contamination of the muon beam  and from cosmic muons. 
The other and more important  background originates from false identifications of muon decays 
as  $\upmu$Au events.
The dark brown (dashed-dotted) curves in  Fig.~\ref{fig:finalHits_c} and  \ref{fig:finalHits_d} represent the integral number of muon decays occurring between  1000~ns and the time $t$. Even in the optimal event time window there is one order of magnitude more muon decays than  laser-induced $\upmu$Au events (note that these curves have been scaled by  a factor of five to ease the representation).
To mitigate this possible background source, the  detection system has to minimise
the false identification of muon decay events as $\upmu$Au events, while maximising the detection probability for $\upmu$Au events. 

A high detection efficiency for $\upmu$Au events is enabled by surrounding the hydrogen target with two clusters of Bismuth Germanium Oxide (BGO) scintillators as shown in Fig.~\ref{fig:target-region} read out by photomultiplier tubes. In total, 18 hexagonal BGO scintillators with a thickness of about 50~mm are arranged up-stream and down-stream of the hydrogen target to allow a large solid-angle coverage. To distinguish the electrons from regular muon decays from $\upmu$Au cascade events, we placed thin (5~mm) plastic scintillators between the BGO clusters and the hydrogen target. This detection scheme was qualified with a prototype including a realistic material budget for the target region.  We obtained a detection efficiency for $\upmu$Au events of $\varepsilon_\text{Au}=0.7$, and a probability of $\varepsilon_\text{Au-false}=0.09$ to falsely identify decay electrons as $\upmu$Au events~\cite{Laura-Thesis}.

\section{Signal and background rates and target optimization}
\label{sec:SigBGRates}

In this section we first summarize how the simulation of the diffusion process can be used to estimate the signal and background rates of the HFS experiment.
The target pressure, temperature and length are then optimized such that the largest statistical significance is reached.

The signal rate $R_\text{signal}$ can be estimated using 
\begin{eqnarray}
    R_\text{signal}& = &P_\text{signal} \cdot R_\upmu\cdot \varepsilon_\text{Au}\\
                   & = & 2.5\times 10^{-4} \cdot 500~\mathrm{s}^{-1}\cdot 0.7= 0.088 ~\mathrm{s}^{-1}\, , \nonumber
\end{eqnarray}
where $R_\upmu=500~\mathrm{s}^{-1}$ is the muon trigger rate in the entrance detector (limited by the laser repetition rate) and $\varepsilon_\text{Au}$ is the detection efficiency for a $\upmu$Au event.
We use here a value of $\varepsilon_\text{Au}=0.7$ as discussed in the end of Sec.~\ref{sec:post-laser}.
$P_\text{signal}$ is the probability that a muon passing the entrance detector eventually produces a $\upmu$Au atom that reaches one of the target walls in the event time window after undergoing a laser-excitation.
This value can be extracted from the orange curves of Fig.~\ref{fig:finalHits_c} and Fig.~\ref{fig:finalHits_d} considering the  difference between beginning and end of the event time window.
For an event  time window spanning from 1.1 to 1.5~$\upmu$s, we obtain $P_\text{signal}=2.5\times 10^{-4}$  when assuming an in-coupled pulse energy of 1~mJ, a cavity reflectivity of \SI{99.2}{\percent}, a laser bandwidth of 100~MHz, and a target with 0.6~bar pressure, 22~K temperature, 1~mm  thickness and 15~mm diameter.

The rate of the  diffusion-related background (see Fig.~\ref{fig:finalHits_a} and Fig.~\ref{fig:finalHits_b})
is given by
\begin{eqnarray}
    R_\text{BG}^\text{diffusion}& = &P_\text{diffusion} \cdot R_\upmu\cdot \varepsilon_\text{Au}\\
                   & = & 11.1\times 10^{-4} \cdot 500~\mathrm{s}^{-1}\cdot 0.7= 0.39~\mathrm{s}^{-1} \, ,\nonumber
\end{eqnarray}
where $P_\text{diffusion} $ is the probability that a muon passing the entrance detector diffuses in the hydrogen gas and hits a target wall in the event time window.
This value can be  extracted from the blue curves of Fig.~\ref{fig:finalHits_c} and Fig.~\ref{fig:finalHits_d} considering the  difference between the beginning and end of the event time window.

The background rate caused by muon decay  is given by 
\begin{eqnarray}
    R_\text{BG}^\text{decay}& = &P_\text{decay} \cdot R_\upmu\cdot \varepsilon_\text{Au-false}\\
                   & = & 67.5\times 10^{-4} \cdot 500~\mathrm{s}^{-1}\cdot 0.09= 0.30~\mathrm{s}^{-1} \, ,\nonumber
\end{eqnarray}
where $P_\text{decay} $ is the probability that a muon entering the target decays in the event time window.
$P_\text{decay} $ can be  extracted from the dark brown (dashed-dotted) curves of Fig.~\ref{fig:finalHits_c} and Fig.~\ref{fig:finalHits_d}.
%
$\varepsilon_\text{Au-false}$ is the probability that an electron from a muon decay is falsely identified in the detection system as a $\upmu$Au event (see Sec.~\ref{sec:post-laser}).

Accidental energy depositions in the event time window are responsible for  a third type of background.
These energy depositions are mainly produced by the natural radioactivity present in the experimental area, and by  neutrons produced by the accelerator complex.
Based on the measured accidental rate in the x-ray detection system~\cite{Laura-Thesis}, we predict an accidental background rate in the spectroscopy experiment of $R_\text{BG}^\text{accidental}=0.2~\mathrm{s}^{-1}$ for a muon rate of $R_\upmu=500~\mathrm{s}^{-1}$ and a  400~ns wide event time window (which determines the duration of the exposure to the accidental background).

The signal and background rates have been computed for various target conditions and the results are summarized in Table~\ref{tab:RateResults}.
The target conditions have been varied to find the maximum of the ratio
$R_\text{signal}/\sqrt{R_\text{BG}}$, where $R_\text{BG}=R_\text{BG}^\text{diffusion}+R_\text{BG}^\text{decay}+R_\text{BG}^\text{accidental}$ is the total background rate.
Maximizing this ratio is equivalent to maximizing the statistical significance and minimizing the
time needed to expose the signal over  the statistical fluctuations of the  background.
Hence, in Table~\ref{tab:RateResults} we also list the average time needed to observe the signal with a $4\sigma$ significance over the background, which is calculated from the rates using
\begin{equation}
t_{4\sigma}=16\frac{R_\text{BG}}{R_\text{signal}^2}.
\end{equation}
In the following, we use this time $t_{4\sigma}$ as a figure of merit to find the optimal target conditions, and to  estimate the maximum time  needed to search for the resonance.
\begin{figure}
\centering
\centering
\includegraphics[width=0.65\linewidth]{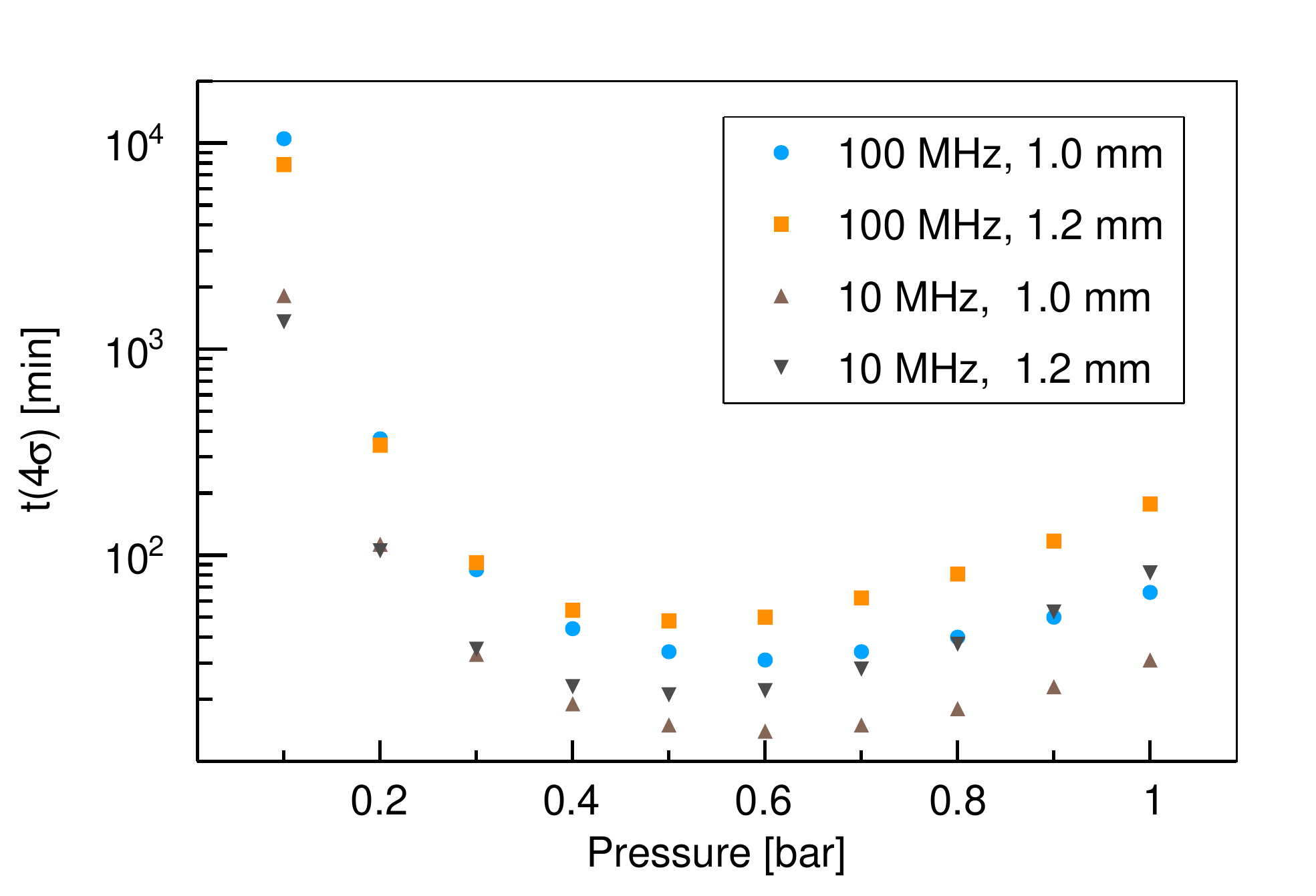}
\caption{Average time $t_{4\sigma}$ needed to observe the signal with a significance over background of $4\sigma$ as a function of the H$_2$ gas pressure for two target thicknesses ($d=1.0~$mm and $d=1.2~$mm and two laser bandwidths (100~MHz and 10~MHz).}
  \label{fig:4sigma-optimization}
\end{figure}
\begin{table}
\caption{\label{tab:RateResults} Signal rates, background rates and $t_{4\sigma}$ for two different pressures $p$, two target thicknesses $d$ and two laser bandwidths $\Delta_l$. We use $R_\upmu=500~\mathrm{s}^{-1}$, $\varepsilon_\text{Au}=0.7$, and $\varepsilon_\text{Au-false}=0.09$. Above the horizontal line we used the $50/50$ distribution of the initial $\upmu$p energy. Below the line we assumed that all the $\upmu$p atoms have an initial kinetic energy of 100~eV. To compute the signal rates we further  assume that the laser is on resonance, with an in-coupled pulse energy of 1.0 mJ, a cavity reflectivity of \SI{99.2}{\percent} and a pulse length of 50~ns.}
\centering
\begin{tabular}{ccccccccc}
 \hline
 $\Delta_l$ & $d$ &  $p$ & $R_\mathrm{Signal}$ & $R^{\mathrm{diffusion}}_\mathrm{BG}$& $R^{\mathrm{decay}}_\mathrm{BG}$& $R^{\mathrm{accidental}}_\mathrm{BG}$ & $R_\mathrm{Signal} / R_\mathrm{BG}$ & $t_{4\sigma}$ \\

  [MHz] & [mm] & [bar] & [1/s] & [1/s]& [1/s] & [1/s] &  & [min]\\
   \hline
100 &  1.0 &  0.5 &  7.7e-02 &  3.2e-01 &  2.2e-01 &  0.2   &   1.0e-01  & 34  \\  
100 &  1.0 &  0.6 &  8.8e-02 &  3.9e-01 &  3.0e-01 &  0.2   &   9.8e-02  & 31  \\  
100 &  1.2 &  0.5 &  7.0e-02 &  3.5e-01 &  3.1e-01 &  0.2   &   8.1e-02  & 48  \\  
100 &  1.2 &  0.6 &  7.4e-02 &  4.1e-01 &  4.1e-01 &  0.2   &   7.3e-02  & 50  \\  
10 &  1.0 &  0.5 &  1.2e-01 &  3.2e-01 &  2.2e-01 &  0.2   &   1.6e-01  & 15  \\  
10 &  1.0 &  0.6 &  1.3e-01 &  3.9e-01 &  3.0e-01 &  0.2   &   1.5e-01  & 14  \\  
10 &  1.2 &  0.5 &  1.1e-01 &  3.5e-01 &  3.1e-01 &  0.2   &   1.2e-01  & 21  \\  
10 &  1.2 &  0.6 &  1.1e-01 &  4.1e-01 &  4.1e-01 &  0.2   &   1.1e-01  & 22  \\  
\hline 
100 &  1.0 &  0.5 &  5.9e-02 &  2.4e-01 &  1.7e-01 & 0.2 & 9.7e-02  & 47  \\
100 &  1.0 &  0.6 &  7.1e-02 &  3.0e-01 &  2.4e-01 & 0.2 & 9.5e-02  & 40  \\
100 &  1.2 &  0.5 &  5.7e-02 &  2.7e-01 &  2.5e-01 & 0.2 & 7.9e-02  & 60  \\
100 &  1.2 &  0.6 &  6.3e-02 &  3.3e-01 &  3.5e-01 & 0.2 & 7.2e-02  & 59  \\ 
10 &  1.0 &  0.5 &  8.9e-02 &  2.4e-01 &  1.7e-01 & 0.2 & 1.5e-01  & 21  \\  
10 &  1.0 &  0.6 &  1.1e-01 &  3.0e-01 &  2.4e-01 & 0.2 & 1.4e-01  & 18  \\  
10 &  1.2 &  0.5 &  8.6e-02 &  2.7e-01 &  2.5e-01 & 0.2 & 1.2e-01  & 26  \\  
10 &  1.2 &  0.6 &  9.6e-02 &  3.3e-01 &  3.5e-01 & 0.2 & 1.1e-01  & 26  \\  
 \hline
\end{tabular}
\end{table}
The pressure dependence of $t_{4\sigma}$ is shown in Fig.~\ref{fig:4sigma-optimization} for a  constant target temperature of 22~K, for two laser bandwidths, and two target thicknesses.
In this parameter range, the minimum is reached for pressures around 0.5~bar, which motivates the pressure choice  in Table~\ref{tab:RateResults}. 
Temperatures lower than 22~K would lead in principle to shorter $t_{4\sigma}$ times (smaller $R_\text{BG}^\text{diffusion}$ and larger laser transition probability), but are excluded to avoid liquefaction of the H$_2$ gas.
Target lengths slightly smaller than 1~mm would also lead to smaller $t_{4\sigma}$, 
but this would also cause larger diffraction losses for the laser light passing the region between the two target walls. This effect cannot be easily quantified as it is strongly dependent on the relative alignment between the laser beam and multi-pass cavity.
For this reason, in this study we assume that the minimal possible target length is $d=1$~mm and for the conservative estimates  we use $d=1.2$~mm.

To study the sensitivity of the results to the initial kinetic energy of the $\upmu$p atoms, in Table~\ref{tab:RateResults} we list rates and $t_{4\sigma}$  values for the two different models of the initial kinetic energy of the $\upmu$p atoms as described in Sec.~\ref{sec:formation}.
Using the most conservative assumption of the  initial kinetic energies (all $\upmu$p at 100~eV energy) instead of the $50/50$ distribution, $t_{4\sigma}$ is increased by 20 to \SI{40}{\percent} depending on the other parameters. 

As can be seen from Table~\ref{tab:RateResults}, for the most conservative estimate of the performance we obtain $t_{4\sigma}\approx 60$~min. 
This can be considerably improved by decreasing the target length or the laser bandwidth. In addition, note that $t_{4\sigma}$ is inversely proportional to the square of the laser energy and the cavity lifetime, such that an improvement of the performance of the laser system and the multi-pass cavity can significantly decrease $t_{4\sigma}$.

\section{Search and scan of the HFS resonance}
\label{sec:resonance}

In this section we first roughly estimate the maximum time needed to search for the resonance and then we calculate the statistical accuracy with which the resonance frequency can be determined.
For both cases, we need to assume a certain performance of the experimental setup and to use the simulation of the diffusion process presented above.

The HFS resonance, which has a linewidth of about 230~MHz at FWHM~\cite{Amaro:2021goz}, has  to be searched for in a region of about 40~GHz  corresponding to a $\pm3\sigma$ band of the present theory uncertainty given mainly by the uncertainty of the two-photon exchange contribution~\cite{Antognini:2022xoo, Peset:2016wjq, Hagelstein:2015lph, Hagelstein:2018bdi, Carlson:2011af, Tomalak:2018uhr, Hagelstein:2015egb, Faustov:2006ve}. 
An efficient search of the resonance has to be accomplished in steps of about 100~MHz corresponding to about 0.5~FWHM,  such that the maximum number of frequency points that needs to be measured is about 400.
A recent evaluation of the two-photon exchange contribution in muonic hydrogen~\cite{Antognini:2022xoo} that uses the two-photon exchange contribution extracted from regular hydrogen has decreased its uncertainty by a factor of two.
However, we still conservatively assume a  40~GHz wide search region.

The resonance will be searched for by counting the number of x rays for a time of $1.4t_{4\sigma}$  at a given laser frequency. 
Then the laser frequency is shifted by 100~MHz and the x-ray counting is performed anew. This is repeated until a statistically significant deviation of the number of x rays above the background level is observed. 
%
%
\begin{figure}
\centering
\includegraphics[width=0.99\linewidth]{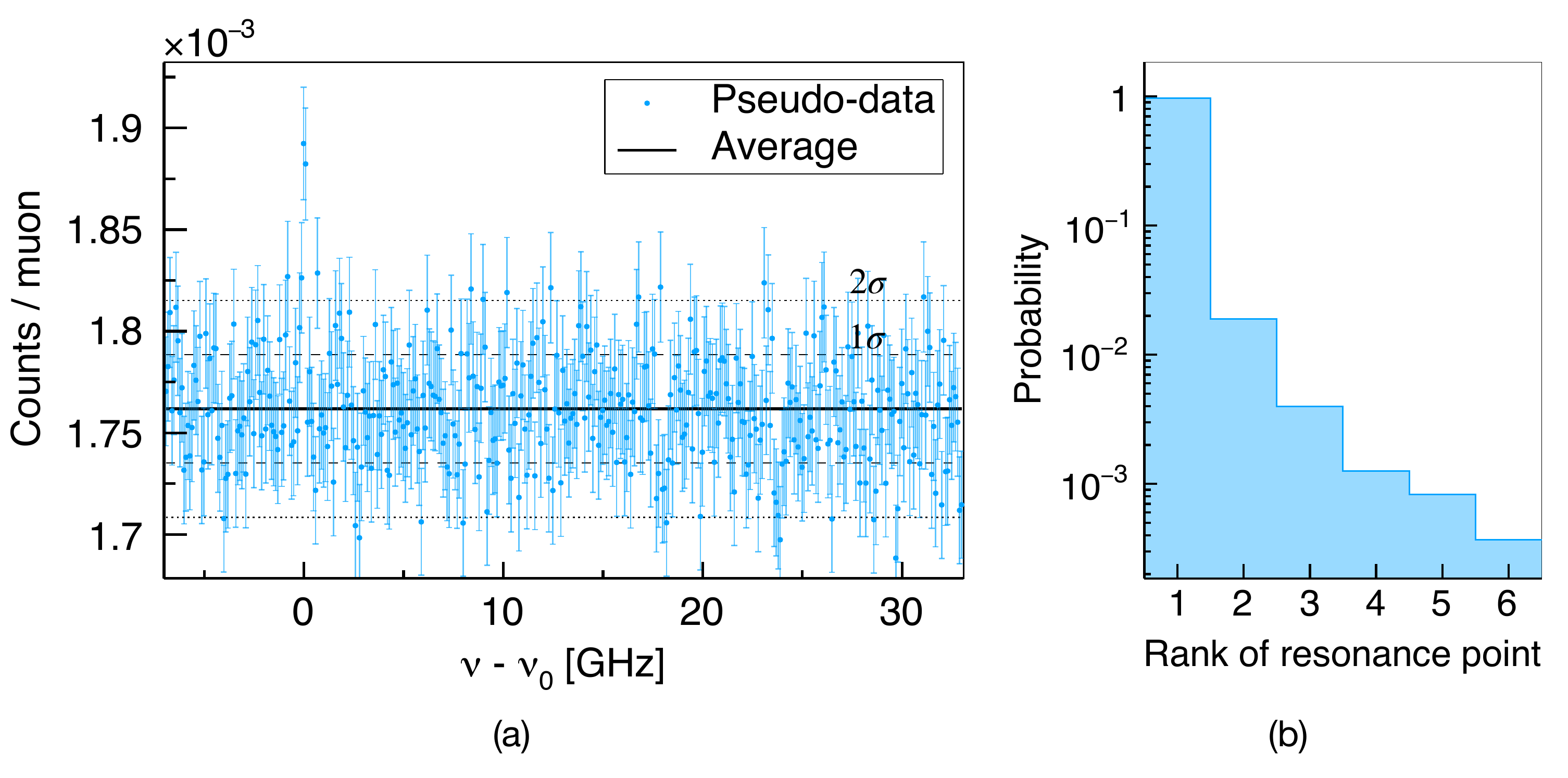}
\caption{(a) Simulation of the resonance search in which a time $1.4t_{4\sigma}$ is spent at each frequency point. $\nu$ denotes the laser frequency, $\nu_0$ the resonance frequency. (b) Ranking of the frequency points in correspondence of the resonance. For \SI{97.43}{\percent} of the cases the simulated pseudo-data have a maximum in  correspondence of the resonance. For \SI{1.84}{\percent} of the cases the second largest point is in correspondence of the resonance and so on. }
\label{fig:resonanceSearch}
\end{figure}
The factor of 1.4 multiplying $t_{4\sigma}$ takes into account that, while searching for the resonance, none of the  chosen frequency points necessarily matches the resonance maximum.
As the maximum deviation from resonance is 50~MHz (for 100~MHz frequency steps) the maximum decrease of the 
laser-induced transition probability is  0.86, resulting in an increase of the time needed to see the signal over background by a factor of $1/0.86^2=1.4$.

This procedure to search for the resonance has been validated by simulating the search for  the resonance $10^5$ times.
The number of events at each frequency point (spaced by 100~MHz) is generated according to a Poissonian distribution with expectation values derived from the signal and background rates given in Table~\ref{tab:RateResults} and accounting for the wavelength dependence  of the signal rate according to Ref.~\cite{Amaro:2021goz}.
One of these simulations is shown in Fig.~\ref{fig:resonanceSearch}a. As visible in this figure, the maximum of the counts has been found at the frequency which is nearest to the resonance position $\nu_0$ used to generate the pseudo-data.

By considering the simulated  resonance searches, we found that for \SI{97.43}{\percent} of the cases, the maximum of the simulated pseudo-data has been found in correspondence of  the resonance (see Fig.~\ref{fig:resonanceSearch}b).
As can be seen from the same figure, the probability that only the second highest point is at the position of the resonance is about \SI{1.84}{\percent} and the probability that only the third highest point is at the position of the resonance is \SI{0.39}{\percent}.
Summing up these probabilities, we obtain \SI{99.66}{\percent}, which corresponds to the probability of identifying the position of the resonance by adding additional measurement time to the three points with the largest amplitude.
Hence, we confirm that the above described simple procedure to search for the resonance with 100~MHz steps and by accumulating statistics at each frequency point for a time of 
$1.4t_{4\sigma}$ is adequate.

The maximum time needed to search for the resonance (using the simple procedure described above) can be estimated to be $400\times (1.4t_{4\sigma}  + t_\mathrm{\lambda-change}) \frac{1}{\varepsilon_\text{uptime}}=82'300$~minutes, corresponding to 8.2 weeks.
For this estimate, we have used conservative values for the experimental performance:  an uptime  (including accelerator) of $\varepsilon_\text{uptime}=\SI{70}{\percent}$, a  time  $t_\mathrm{\lambda-change}=1$~h to change the laser frequency, a laser pulse energy of 1~mJ, a laser bandwidth of 100~MHz, a cavity reflectivity of \SI{99.2}{\percent}, a muon rate of $500~\mathrm{s}^{-1}$, $\varepsilon_\text{Au}=0.7$, $\varepsilon_\text{Au-false}=0.09$, a target thickness of 1.2~mm, and scan range of 40~GHz. Moreover, we assumed that all $\upmu$p atoms have 100~eV initial kinetic energy. 
Most probably the resonance can be found much faster if a significant deviation from background is found earlier and by adapting the search strategy (i.e. accumulating more statistics on points with significant deviations from background).

We have also simulated $10^5$ pseudo-measurements of the HFS resonance after its discovery, assuming that two weeks of beamtime can be used to measure the resonance (this time does not include the time needed to change the laser wavelength and the time when the setup or the accelerator are not operating).
\begin{figure}
\centering
\begin{subfigure}{0.49\textwidth}
\centering
\includegraphics[width=0.99\linewidth]{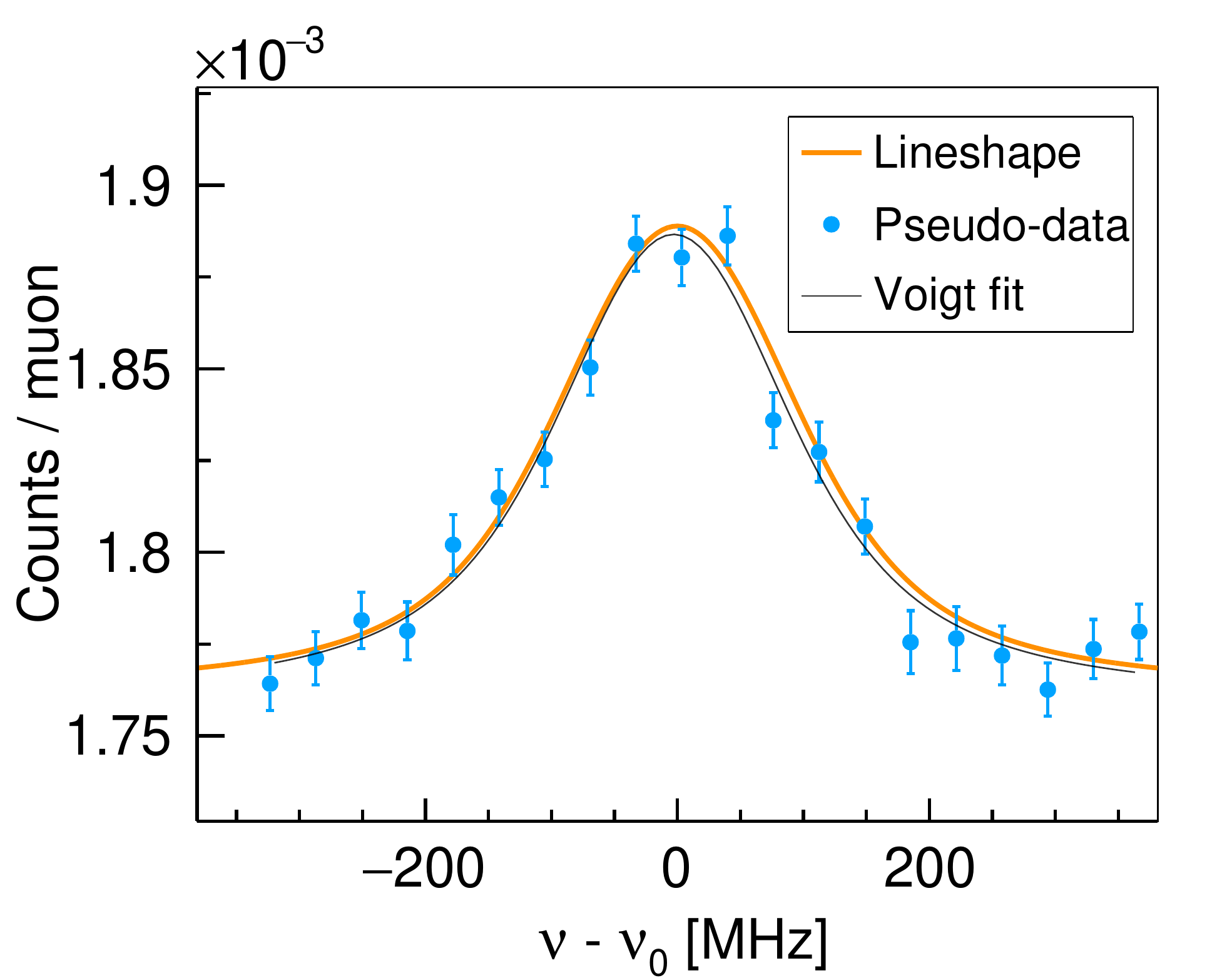}
\caption{}
\label{fig:resonancePlot_a}
\end{subfigure}
\begin{subfigure}{0.49\textwidth}
\centering
\includegraphics[width=0.99\linewidth]{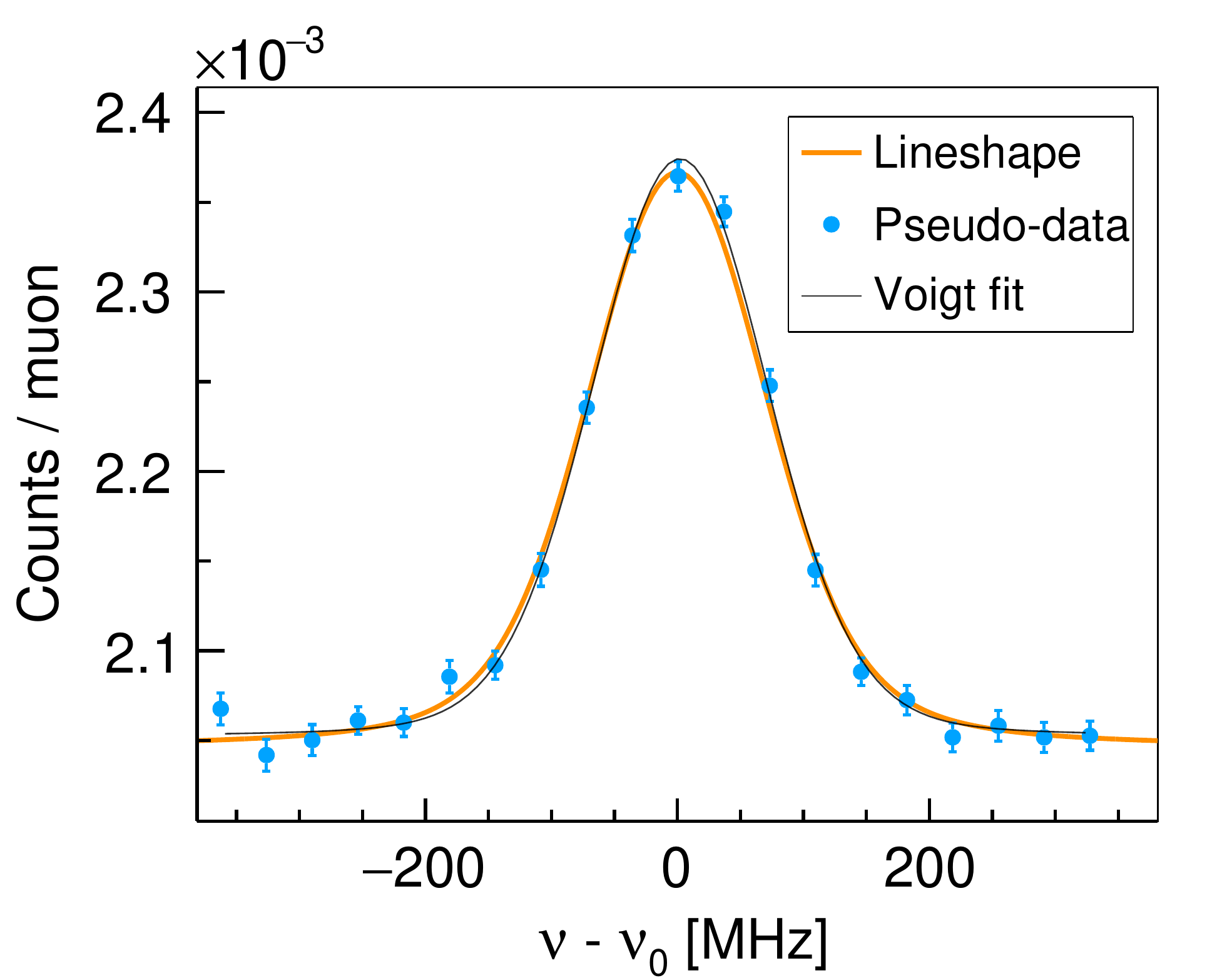}
\caption{}
\label{fig:resonancePlot_b}
\end{subfigure}
\begin{subfigure}{0.49\textwidth}
\centering
\includegraphics[width=0.99\linewidth]{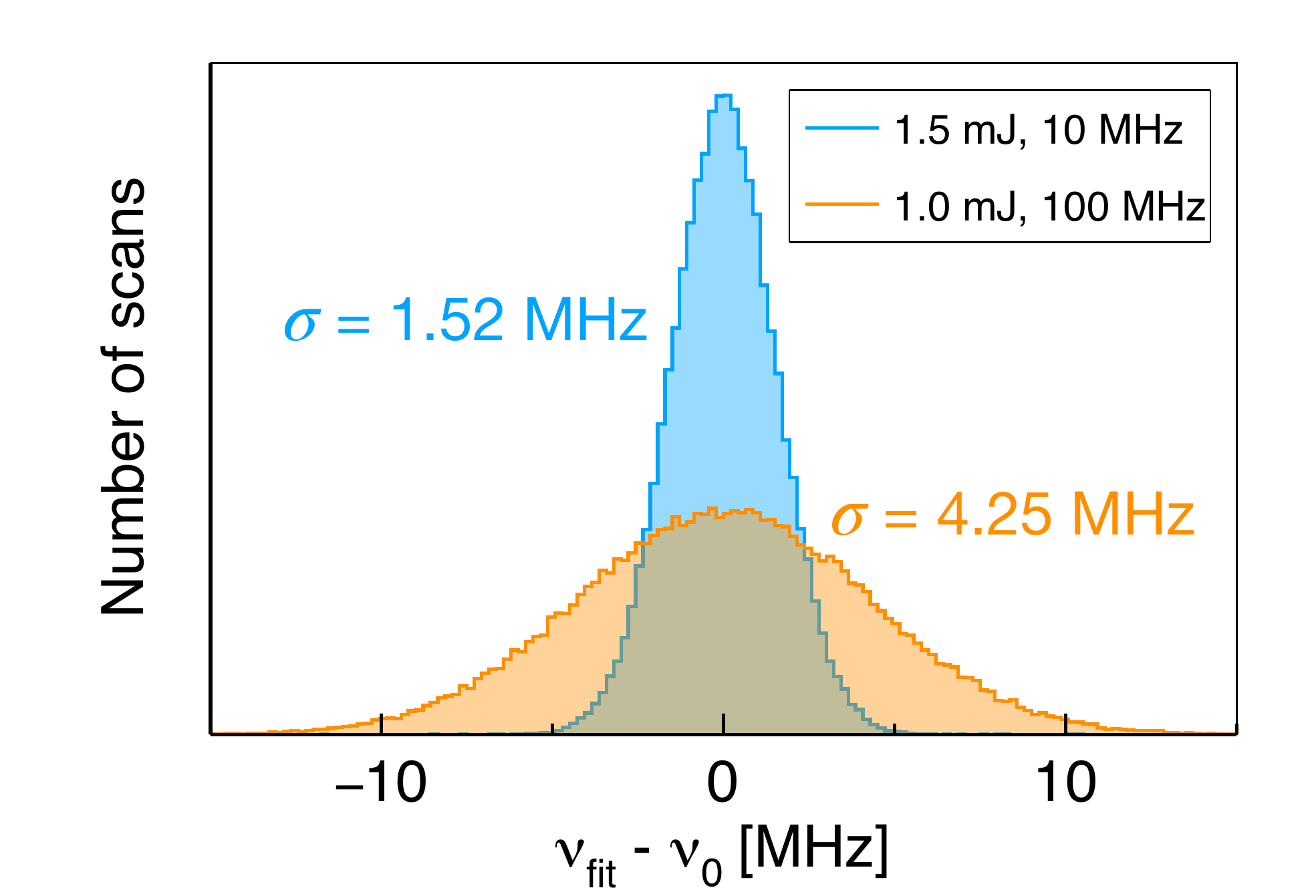}
\caption{}
\label{fig:resonancePlot_c}
\end{subfigure}
\begin{subfigure}{0.49\textwidth}
\centering
\includegraphics[width=0.99\linewidth]{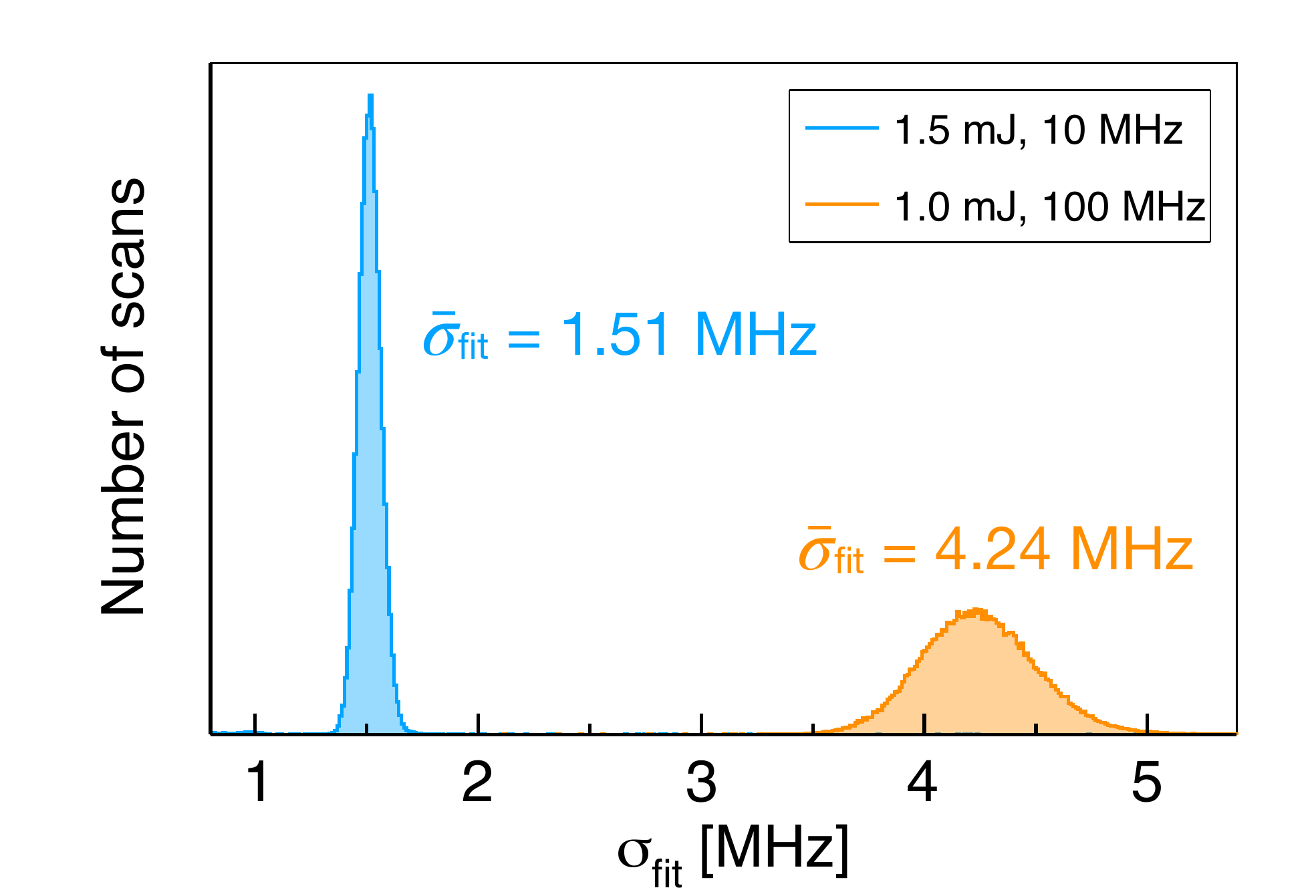}
\caption{}
\label{fig:resonancePlot_d}
\end{subfigure}
\caption{(a) Simulated pseudo-measurement of the HFS transition for two weeks of measurement. We assumed here an initial $\upmu$p energy of 100~eV, a laser pulse energy of 1~mJ, a laser bandwidth of 100~MHz, a cavity reflectivity of \SI{99.2}{\percent}, a muon rate of $500~\mathrm{s}^{-1}$, $\varepsilon_\text{Au}=0.7$, $\varepsilon_\text{Au-false}=0.09$, a target thickness of 1.2~mm and an accidental rate of $0.2~\mathrm{s}^{-1}$. (b) Similar to (a) but for a pulse energy of 1.5~mJ, a laser bandwidth of 10~MHz and the more realistic initial kinetic energy distribution as presented in Sec.~\ref{sec:formation}. (c) Distribution of the bias obtained by fitting a Voigt function to the resonance data for $10^5$ pseudo-measurements. The orange distribution has been obtained  from data generated at the conditions used in (a), the blue distribution from data generated at the conditions used in (b). (d) Fit uncertainties of the centroid position obtained from the fits discussed in (c). }
\label{fig:resonancePlot}
\end{figure}
A simulation of a resonance measurement for conservative assumptions is shown in Fig.~\ref{fig:resonancePlot_a}. 
 Figure~\ref{fig:resonancePlot_b} shows a similar simulation  for slightly less conservative assumptions  to highlight the potential of the experimental scheme.
In both cases we have chosen regular frequency steps of 35~MHz and we have distributed a similar amount of time to each frequency point with maximum fluctuation of about \SI{20}{\percent}.
Hence, we did not put any effort into optimizing the distribution of the available time at the various laser frequencies.

The obtained pseudo-experiments were then fit using a Voigt function to determine the  centroid positions.
The results from the fits are given in Figs.~\ref{fig:resonancePlot_c} and ~\ref{fig:resonancePlot_d}.
Figure~\ref{fig:resonancePlot_c}  shows the bias, i.e., the difference between fitted positions and actual position  assumed when generating the pseudo-data. Its distribution is Gaussian centered at zero with  $\sigma=4.25$~MHz and $\sigma=1.52$~MHz, for the most conservative and less conservative assumptions of the experimental performance used to generate Figs.~\ref{fig:resonancePlot_a} and ~\ref{fig:resonancePlot_b}, respectively. 
Figure~\ref{fig:resonancePlot_d} gives the distributions of the fit uncertainties of the centroid position. The results are equivalent to the standard deviation of the bias, demonstrating that fitting with a simple  Voigt profile is adequate.

From these plots, we can thus conclude that a statistical  uncertainty  of a few  MHz can be reached  for the HFS spectroscopy experiment even with the most conservative assumptions of the setup performance. This corresponds to a relative accuracy of about 0.1~ppm to be compared with the goal of 1~ppm.
Hence, even assuming  
a conservative  performance of the experimental setup, it should be possible to find and measure  the resonance  with relative accuracies below the ppm level within a beamtime having a duration less than 12 weeks which is approximately the maximum time that can be allocated at the PSI CHRISP facility.

\section{Conclusion}
\label{sec:conclusion}

In this paper we presented simulations of the diffusion of $\upmu$p atoms in the hydrogen gas
prior and after the laser excitation to optimize  the hydrogen target conditions, to estimate signal and background rates and to asses the statistical uncertainty of the HFS experiment.
The simulations were implemented into Geant4 using double-differential collision rates for the scattering between $\upmu$p in the ground state and H$_2$ molecules. Possible excitation of the H$_2$ molecules and transitions  of the $\upmu$p atom between the singlet and triplet sub-levels are accounted for.
Because the kinetic energy of the hydrogen molecules can not be neglected compared with the kinetic energy of the $\upmu$p atoms, we accounted also for the initial kinetic energy of the H$_2$ molecules.
This has been obtained by calculating the double differential collision rates averaged over the distribution of velocities of the hydrogen molecules as given in Eq.~(\ref{eq:rates_lab}).
In this way, 
the Geant4  simulations were capable to include, in an effective way, the H$_2$ motion and thus the gas target temperature.
We applied a similar code with similar cross sections to quantify the diffusion of muonic deuterium in a high pressure H$_2$ gas cell in the context of the muX experiment and we found good agreement between simulation and measurements~\cite{muX:2022}.

In the presented simulations we have  used a conservative approximation of the
initial kinetic energy (after the $\upmu$p atom formation and deexcitation) derived from the results of the cascade model presented in Ref.~\cite{Covita:2017gxr} for $\varphi=0.015$ and modified to account for the fact that the $\upmu$p-H$_2$ cross sections are available only up to 100~eV.
By considering the extreme case that all $\upmu$p atoms after the deexcitation would have 100~eV kinetic energy, we 
have investigated the sensitivity of the results to the initial kinetic energy and derived lower bounds for the signal and background rates.

With the diffusion simulations we computed the processes that lead to a signal event and the two main backgrounds.
Combined with the measured values for the detection efficiencies $\varepsilon_\text{Au}=0.7$ and $\varepsilon_\text{Au-false}=0.09$,
we were  able to estimate  signal and background rates.
We optimized the target conditions using the ratio $R_\text{signal}/\sqrt{R_\text{BG}}$ as a figure of merit and we found that the optimal pressure is between 0.5 and 0.6 bar for target temperatures of 22~K (corresponding to $\varphi= 0.008-0.01$) and target thicknesses between 1 and 1.2 mm.
Lower temperatures are not possible as the hydrogen gas liquefies, while target lengths smaller than 1~mm would lead to diffraction losses of the laser beam.

Using these optimized target conditions (0.6 bar and 22 K) and conservative estimates of the other still unknown parameters (1 mJ in-coupled laser pulse energy, a cavity reflectivity of \SI{99.2}{\percent}, a laser bandwidth of 100~MHz and 1.2 mm target thickness) we have estimated the maximum time needed to search for the resonance.
Nine weeks of beam time should be sufficient to find the resonance even when the resonance has to be searched in a 40~GHz wide region.
With additional two weeks of data taking (corresponding to  three weeks of beam time), it is possible to measure the HFS transition with sub ppm statistical uncertainty.
Turning around the point of view, these simulations allow to precisely define the  requirements for the experimental setup.

%
\section*{Acknowledgments}

We acknowledge the support of the following grants: FCT - Funda\c{c}\~{a}o
para a Ci\^{e}ncia e a Tecnologia (Portugal) through national funds in the frame of projects PTDC/FIS-AQM/29611/2017 and UID/04559/2020 (LIBPhys);
Deutsche Forschungsgemeinschaft (DFG, German Research Foundation) under
Germany's Excellence Initiative EXC 1098 PRISMA (194673446);
Excellence Strategy EXC PRISMA+ (390831469) and DFG/ANR Project LASIMUS (DFG Grant Agreement 407008443); 
The French National Research Agency with project ANR-18-CE92-0030-02;
The PESSOA Huber Curien Program 2022, Number 47863UE;
The European Research Council (ERC) through CoG. \#725039;
and the Swiss National Science Foundation through the projects SNF 200021\_165854 and SNF 200020\_197052.


\bibliography{Diffusion_muonicHydrogen_HFS}



\appendix

\end{document}